\documentclass[sigconf]{acmart}

\AtBeginDocument{%
 \providecommand\BibTeX{{%
  \normalfont B\kern-0.5em{\scshape i\kern-0.25em b}\kern-0.8em\TeX}}}

\copyrightyear{2024} 
\acmYear{2024} 
\setcopyright{rightsretained} 
\acmConference[DIS '24]{Designing Interactive Systems Conference}{July 1--5, 2024}{IT University of Copenhagen, Denmark}
\acmBooktitle{Designing Interactive Systems Conference (DIS '24), July 1--5, 2024, IT University of Copenhagen, Denmark}\acmDOI{10.1145/3643834.3660677}
\acmISBN{979-8-4007-0583-0/24/07}

\usepackage[T1]{fontenc}
\usepackage{graphicx}

\begin{document}

\title[Responding to gAI with RtD]{Responding to Generative AI Technologies with Research-through-Design: The Ryelands AI Lab \\ as an Exploratory Study}

\author{Jesse Josua Benjamin}
\orcid{0000-0003-3391-3060}
\email{j.j.benjamin@lancaster.ac.uk}
\affiliation{%
 \mbox{\institution{Lancaster University}
 \country{United Kingdom}}
}

\author{Joseph Lindley}
\orcid{0000-0002-5527-3028}
\affiliation{%
 \institution{Lancaster University}
 \country{United Kingdom}
}
\author{Elizabeth Edwards}
\orcid{0000-0001-8799-0905}
\affiliation{%
 \institution{Lancaster University}
 \country{United Kingdom}
}
\author{Elisa Rubegni}
\orcid{0000-0002-6375-7604}
\affiliation{%
 \institution{Lancaster University}
 \country{United Kingdom}
}



\author{Tim Korjakow}
\orcid{0000-0001-8209-4079}
\affiliation{%
 \institution{Technische Universit\"at Berlin}
 \city{Berlin}
 \country{Germany}
}

\author{David Grist}
\orcid{0009-0001-0658-7469}
\author{Rhiannon Sharkey}
\orcid{0009-0007-8688-2321}
\affiliation{%
 \institution{Ryelands Primary and Nursery School}
 \city{Lancaster}
 \country{United Kingdom}
}

\renewcommand{\shortauthors}{Benjamin, et al.}
\begin{abstract}
Generative AI technologies demand new practical and critical competencies, which call on design to respond to and foster these. We present an exploratory study guided by Research-through-Design, in which we partnered with a primary school to develop a constructionist curriculum centered on students interacting with a generative AI technology. We provide a detailed account of the design of and outputs from the curriculum and learning materials, finding centrally that the reflexive and prolonged `hands-on' approach led to a co-development of students' practical and critical competencies. From the study, we contribute guidance for designing constructionist approaches to generative AI technology education; further arguing to do so with `critical responsivity.' We then discuss how HCI researchers may leverage constructionist strategies in designing interactions with generative AI technologies; and suggest that Research-through-Design can play an important role as a `rapid response methodology' capable of reacting to fast-evolving, disruptive technologies such as generative AI.
\end{abstract}



\received{8 February 2024}
\received[revised]{3 May 2024}


\maketitle

\section{Introduction}
Generative AI technologies have rapidly entered public discourse, awareness and usage since late 2022. ChatGPT was the fastest computer system in history to reach 100 million monthly users \cite{hu_chatgpt_2023}, and the release of Stable Diffusion was the first time such a model was `openly' published \cite{wiggers_stability_2023}. Notable examples of these technologies now residing firmly in the public sphere include widely-shared `fake' images such as the ``Balenciaga Pope,''\footnote{\url{https://www.forbes.com/sites/danidiplacido/2023/03/27/why-did-balenciaga-pope-go-viral/}, accessed 10/11/2023.} the Writers' Guild of America strikes \cite{noauthor_wga_2023}, and governmental reports such as the UK House of Commons committee report stressing the balance between creators' rights and innovation \cite{culture_media_and_sport_committee_connected_2023}. For the HCI community, and design in particular, generative AI technologies have engendered a flood of publications in contexts of use ranging from their integration into design tools or end-user facing components, variously prompting enthusiasm (see e.g. \cite{jiang_promptmaker_2022}) as well as critique (see e.g. \cite{mirsky_creation_2021}). In both the academic and public sphere, uncertainties around the impact of these technologies abound.

In this paper, we adopt Research-through-Design (RtD) as an overarching methodology as one way to respond to these challenges, testing the approach through an exploratory study in a specific context. In contrast to many canonical examples of RtD that are focused on specific artefacts or devices (e.g., \cite{gaver_drift_2004}), we embraced RtD's methodological attributes and qualities including valuing emergence \cite{gaver_emergence_2022} and the aspiration to produce ``intermediate-level knowledge'' \cite{hook_strong_2012} as a primary consideration. In epistemological terms, this necessarily required that we accept our findings would be contingent, and most likely of use for other researchers in a generative rather than prescriptive sense (see \cite{gaver_what_2012}). The practical consequence of these epistemological commitments for our contributions is that they should be seen not as proposals for generalizable theory, but rather as one of many examples which will contribute to an emerging ``research program'' \cite{redstrom_making_2017} of research relating to the role and use of design in the ongoing adoption of generative AI technologies. 

The particular context for testing our approach in an exploratory study is education, as here the adoption of generative AI technologies has caused and is causing significant debates and uncertainties. Text synthesis models have raised fears that students would simply employ such models to `cheat' (e.g., ~\cite{klein_chatgpt_2023}), while technologies intended to detect such usage have been proven to be unable to do so accurately.\footnote{\url{https://openai.com/blog/new-ai-classifier-for-indicating-ai-written-text}, accessed 11/08/2023.} Other avenues include the integration of AI technologies for prediction of educational `performance,' which has garnered extensive criticism (e.g., \cite{perrotta_deep_2020}). Further, for many young people growing up today generative AI technologies will be something they interact with throughout their lives. Hence, there is an emerging consensus (see \cite{department_for_education_generative_2023,giannini_generative_2023}) that considered education strategies are needed to help current and future generations of young people be able to acquire practical and critical competencies for them---in other words, AI literacy (see \cite{long_what_2020}). In this regard, there is a rich tradition in HCI research in educational contexts that seeks to develop literacies for new technologies; for instance regarding ``critical data literacies'' \cite{hautea_youth_2017}, and there is nuanced work on AI technologies coming to the fore (e.g., \cite{judd_all_2020, walsh_making_2022,lee_developing_2021}). However, we observe that opportunities for engaging directly with \textit{actual} AI technologies in a prolonged manner have been limited. In turn, we argue that RtD, with its emphasis on novel materials and extended experiences thereof, can provide opportunities for exploration how young people encounter and form competencies with generative AI technologies.

Building methodologically on RtD, we present an exploratory study that designed and delivered an introductory, \textit{constructionist} curriculum to generative AI technology. Constructionist learning posits that models of the world are built through the direct engagement with \textit{materials} (see \cite{papert_situating_1991,sawyer_constructionism_2005}), and that this is an intrinsic part of how we gain knowledge. In the case of our study this meant that the students were actually \textit{using} the generative AI technology they were learning about, i.e., it was a prolonged, reflective and `hands on' approach to learning. Regarding our overall approach, this means that while we did design several artefacts as part of the project's delivery (e.g., image generation tool, learning materials and printed outputs), the `design' that is at the center of our use of RtD is the design of the curriculum as a whole. In turn, we argue that this offers intermediate-level knowledge both within the context of the study, i.e. generative AI technology education, as well as methodological impetus on the use of RtD to engage with generative AI technologies reflectively. 

The paper is structured as follows. Initially we provide a brief background section discussing related HCI research, focusing on education and AI technologies to outline our motivation for employing RtD. Next, we discuss the RtD approach in more detail, explaining the rationale for and implications of RtD as the overarching methodological framing for this work. For the exploratory study itself, we subsequently describe the reflexive design, development and delivery of a six week curriculum and the production of several post-hoc artefacts and activities, presenting the work in terms of three key phases. In the study specific findings, we document centrally how practical and critical competencies of students appeared to \textit{co}-evolve and report on the teachers' views of the curriculum; and gather propositions for HCI research to engage generative AI education in the form of (1) guidance for designing constructionist curricula on generative AI technologies and (2) reflection on the need for `critical responsivity' due to the potential impact of generative AI technologies in educational contexts.\footnote{To further make our work actionable for the field, we provide an OSF repository with all project materials (documentation images, learning materials, design artefacts, anonymized interview transcripts, project outputs) at \url{https://osf.io/9afnr/?view_only=93c8d0e2159847dba2c454519682b85d}, accessed 24/04/2024.} In our discussion, we then center on the implications stemming from our use of RtD to engage generative AI technologies. Here, we (3) discuss how the design of interactions with generative AI systems may leverage constructionist strategies, and (4) reflect on RtD as a `rapid response methodology' for producing insights pertaining to rapidly evolving disruptive technologies such as generative AI. With these contributions, we aim for HCI researchers to find practical, strategical and methodological support for employing design in the uncertainties characterizing the ongoing adoption of generative AI technologies. 

\section{Related Work}
A full overview of HCI education discourse is beyond the scope of this paper, hence it is important to clarify that we do not refer to education on HCI, but rather HCI \textit{for} education~\cite{pittarello_hci_2017}. There are many AI-based resources for education (for example see this collection \cite{nerantzi_101_2023}), however, education-focused HCI research has much to offer in terms of developing accessible tools for teachers and students. While there are many perspectives one could explore this potential from, in this section we initially focus on the constructionist learning theory which is at the heart of this research before summarizing popular approaches to AI literacy in schools.

Among many paradigms, HCI education research has long built on the learning theory of constructionism as proposed by Papert \cite{papert_situating_1991}. Closely related to constructivism, which tends to focus on learner experience, constructionism posits that learning can arise from active creation, the development of socially meaningful artifacts, and the combination of self-reflection and interaction with others \cite{dangol_constructionist_2023}. The constructionist paradigm explicitly casts students as active learners taking part in interactions with materials which are crucial to learning and knowledge-acquisition (see \cite{banchi_many_2008,kafai_constructionist_2015, levin_constructionist_2017}). The value of constructionism is evidenced in a plethora of toolkits that HCI researchers have developed for researchers, practitioners, students and other stakeholders (see e.g., \cite{blikstein_gears_2013}), and has also served as the guiding philosophy for such hugely successful educational software as \textit{Scratch} \cite{maloney_scratch_2010}. Further, HCI sub-fields such as child-computer interaction have drawn extensively on constructionist principles to combine practical skills and ethics (see \cite{van_mechelen_18_2020}), leading to teaching approaches that reflect a critical stance on technology (e.g., ``critical data literacies'' \cite{hautea_youth_2017}).

There is a high likelihood that generative AI will have a transformative effect on a whole range of social, cultural, and economic issues (see \cite{giannini_generative_2023}). HCI scholarship can offer a valuable perspective by exploring how to bolster AI literacy with younger stakeholders. The UNESCO K-12 AI education framework defines AI literacy as ``some level of competency with regard to AI, including knowledge, understanding, skills, and value orientation'' \cite{unesco_k-12_2022}. The consequence of this framing is that AI literacy transcends declarative or explicit technical knowledge alone and must involve a broader awareness of the technology's societal implications and an ability to critically reflect on these. This aligns with HCI perspectives such as Long and Magerko’s assertion that AI literacy is ``a set of competencies that enables individuals to critically evaluate AI technologies; communicate and collaborate effectively with AI; and use AI as a tool online, at home, and in the workplace'' \cite{long_what_2020}. 

There is a wide range of HCI research related to education and AI, as shown by Su and Zhong in an extensive review \cite{su_artificial_2022}. AI literacy work for middle and high school students includes project-based \cite{judd_all_2020}, art-based \cite{walsh_making_2022} and ethical impact-focused work \cite{schaper_computational_2022, smith_research_2023} as well as more quantitatively oriented work \cite{lee_prompt_2023}. Druga and colleague's work employs co-design and prototyping methods to embody an AI agent or prototype potential systems with primary school children \cite{druga_inclusive_2019}. Williams and colleagues note the importance of younger (pre-school) learners gaining AI literacy, citing safety as a motivation as they encounter AI technologies in aspects of everyday life \cite{williams_popbots_2019}, while Han and Cai explore the benefits and challenges of generative AI through interviews with experts (parents, teachers, technologists) discussing an AI-infused storytelling app \cite{han_design_2023}. 

There have also already been fruitful examples of HCI design research being used to interrogate AI technologies. An exemplary approach is Dove and Fayard’s work on using a metaphorical, playful comparison of Machine Learning (ML) to monsters such as Frankenstein \cite{dove_monsters_2020}, unfolding concerns, worries, and potential mitigation of adverse effects of ML deployments concerning student mental health. Bilstrup and colleagues used RtD to develop an educational tool for creating ML models\cite{bilstrup_opportunities_2022}, and Lindley and colleagues' employed RtD to interrogate ``the emerging reality of living with AI'' \cite{lindley_researching_2020} by designing speculative digital signage. Tamashiro's use of ``design futuring'' to explore contemporary AI technologies \cite{aki_tamashiro_how_2021}, and Rubegni and colleagues' investigation into the hopes and fears children have relating to AI using scenarios \cite{rubegni_dont_2022}, further demonstrate the value of design-led approaches in this context. 

This is clearly a growing and diverse area of research. There are, however, only limited approaches that both involve education practitioners in the process and also center on actual use of the technology as opposed to hypothetical use \cite{williamson_historical_2020}. Further, exploratory work such as Lee and colleagues' ``Prompt aloud!'' \cite{lee_prompt_2023} has yet to conduct a prolonged engagement of learners on a more qualitative footing. A constructionist approach to learning how to use AI technologies may help address the latter limitation by leveraging a direct and sustained engagement with the technology. It is here that we see a promising opening for testing RtD in the generative AI education space: design-led inquiries of actual technologies have the potential to find radical new applications, concepts and frameworks, which the disruptive and practically as well as ethically challenging development of generative AI technology calls for. In the next section, we outline our methodological choices before presenting the Ryelands AI Lab project.

\begin{figure*}
  \centering
  \includegraphics[width=0.9\textwidth]{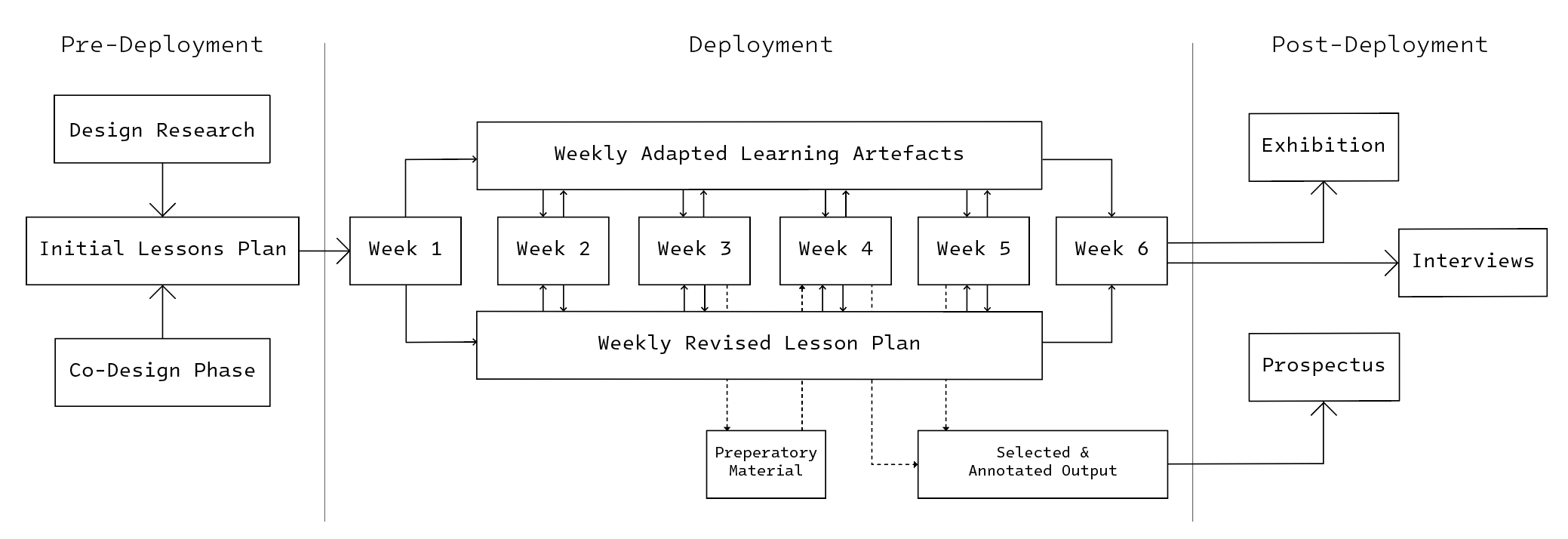}
  \caption{Overview of the exploratory study phases as they manifested \textit{through} the RtD methodology in hindsight.}
  \Description{A diagram showing the exploratory study phases as they manifested through the RtD methodology in hindsight. It is separated into three phases, from left (beginning) to right (end). Left: Pre-deployment Phase. Design Research and a Co-Design Phase feed into the Initial Lesson Plan for the six-week curriculum intervention. Middle: Deployment Phase. Six weekly lessons are linked through feedback loops to Weekly Adapted Learning Artefacts and Weekly Revised Lesson Plan. Between Week 3 and Week 4, a further feedback look integrates Preparatory Material. Week 4 and Week 5 feed into Selected and Annotated Output, which connects to subsequent phase, and Week 6 connects into the subsequent phase separately. Right: Post-Deployment Phase. Week 6 feeds into the Exhibition and Interviews with teachers, while the Selected and Annotated Output feeds into the Prospectus.}
  \label{fig:phases}
\end{figure*}

\section{Methodology}
In this section, we detail the methodological approach behind the Ryelands AI Lab, a constructionist curriculum intervention spanning six weekly lessons with students aged 7 to 9 and a subsequent exhibition of outputs at Ryelands Primary and Nursery School in Northwest England. We also provide contextual detail and provide a consideration of our positionality.

\subsection{Research-through-Design as Guiding Methodology}
Within our exploratory study, our primary objective was to explore the process of delivering a constructionist approach to teaching practical and critical competencies for generative AI technologies in a primary education context. Such competencies form, according to Long and Magerko, the core of AI literacy: understood not as purely declarative knowledge (i.e., `AI is X') that one receives but rather the kind of knowledge that promotes reflection as well as practical use \cite{long_what_2020}. To support this kind of AI literacy required the design and development of brand-new learning materials and resources. This is one reason for deciding to employ Research-through-Design. RtD is a methodology that, in general, helps ``researchers to investigate the speculative future, probing on what the world could and should be'' \cite{stappers_research_2019} through the design-led exploration of materials, scenarios and concepts. This is achieved via the process of designing objects, services and other artefacts and leveraging both reflective practice during the process as well as the resultant outcomes to discuss and explore potential impact and implications (see \cite{stappers_research_2019,gaver_what_2012,bardzell_immodest_2015,lindley_implications_2017,zimmerman_research_2007,zimmerman_analysis_2010}). As highlighted by Williamson, this is an important matter in the generative AI education space insofar as that methods are needed with which to create ``alternative social laboratories'' \cite{williamson_degenerative_2023} that can counter the industry dominance of how these technologies enter society. Coupled with our observation that studies of prolonged engagement of young people with actual generative AI technologies are lacking, this forms the basis for our rationale to engage this context with RtD: The reflective and responsive design \textit{and} simultaneous study of a curriculum intervention as a `lab' for hands-on engagement with adapting materials is made methodologically possible through RtD, offering insights that are not driven by industry headlines or public fears but rather center on how young people engage this disruptive technology directly.

RtD can be a predominantly ``artefact-centric'' \cite{gaver_what_2012} \textit{method}, where it is a ``thing-making practice whose objects can offer a critique of the present and reveal alternative futures'' \cite{bardzell_immodest_2015}. Our work, however, adopted RtD as a guiding \textit{methodology}. This allowed us to lean on one of RtD's main epistemological strengths; its reflexivity \cite{sengers_staying_2006}, meaning the capacity to engage with a domain or material while continuously ``reframing the underlying situation and goal during the design process'' \cite{olson_research_2014}. An alternative acknowledgement of the requirement to be flexible about the RtD process is framed as ``drifting by intention'' \cite{krogh_drifting_2020}. As a result, in RtD knowledge is not expected to emerge as a ``bounded thing'' \cite{stappers_research_2019} related to one artefact or hypothesis, but rather as taking shape alongside the various activities of \textit{doing} RtD. In other words, RtD here did not mean that we planned to design one specific artefact and study how its use and contextual embedding would prompt certain propositions or questions. Rather, and similar to epistemological considerations in participatory design (see \cite{halskov_diversity_2015,schuler_participatory_1993}), we sought to use design to develop a holistic epistemological position on the entirety of a particular context: the design and delivery of a constructionist curriculum on generative AI technologies.

It would have been feasible to imagine adopting a different research methodology that involved creating a curriculum and learning materials, and then studying the efficacy of those externally, for example by establishing a control group and a study group and comparing the impact of the curriculum. Such an approach certainly would have some virtues, however, to facilitate a rapid response to the newly-released technology (in our case we were working with Stable Diffusion which was released only months prior), RtD allowed us to combine the process of designing the curriculum while also assessing its value. An implication of this approach is that the insights we would create would be be a type of ``intermediate knowledge'' \cite{hook_strong_2012}. This imports some limitations, for example our findings are based on sample of one (i.e., only one instance of the constructionist curriculum was evaluated). Balancing this limitation is the acceptance that the findings are intended mainly to inform future work, contributing to an ongoing program of research, as opposed to positing that a hypothesis has been evidenced beyond reasonable doubt. Lastly, we were also driven by wanting to better understand how RtD can methodologically respond to a disruptive technology in a specific context---here, generative AI technology in education. 

The diagram we present in Figure \ref{fig:phases} is a post-hoc representation of how the exploratory study was structured. Elements of this were planned, although some of the structure emerged across the process. The deployment phase depicted was a particularly intense period for the project and involved weekly adaptations to our learning materials and artefacts, ensuring that our learning that developed on a week-by-week basis was incorporated into the subsequent weeks. In the context of this project, this reflexive quality was a clear benefit of the RtD methodology. Our starting point was undoubtedly provisional, but this allowed us to adapt and shape our findings to provide as holistic a point of view as possible, while accepting the limits of intermediate knowledge (we reflect on further nuances of this process in section \ref{ssec:RtDRRM}). 

\subsection{Contextual Overview}
Here, we document the specific demands and aspects of engaging in the chosen context of primary school education, and how specific aspects such as recruitment and analysis were shaped by our RtD methodology.

\subsubsection{Recruitment and Approach to Context}\label{ssec:approach}
Prior to the exploratory study, the Head Teacher of the school we were working with suggested working with their year 4 students (aged 7-9). This suggestion was based partly on the considerations that the research intervention would not interfere with any national exams. Additionally, in contrast to younger students this group generally have the skills to use their computers without assistance. After a briefing on the goals and technology involved, we began regular meetings with two year 4 teachers. Each of the teachers has a year 4 class of 25 students, and the teacher deals with all aspects of the curriculum and delivers all lessons for the class they are responsible for. 

\subsubsection{Demographics}
Our participants were preselected as those who were in the two year 4 classes at the school we were working with. The school is situated in an economically and socially disadvantaged area (see section \ref{ssec:interviews}). This is also reflected in the high share of students' school meals supported by UK government pupil premium grants,\footnote{\url{https://www.gov.uk/government/publications/pupil-premium/pupil-premium}, accessed 08/08/2023.} at 63\% of the student population.\footnote{\url{https://ryelands.lancs.sch.uk/pupil-premium/}, accessed 24/04/2024.} This makes it roughly representative of the North-West England county it is situated in as a whole, whose GDP sits at 67.7\% of the UK national average, and which is also ethnically predominantly White-British.\footnote{\url{https://ryelands.lancs.sch.uk/wp-content/uploads/2022/11/Single-Equalities-Policy-May-2022.pdf}, accessed 24/04/2024.} 

\subsubsection{Research Ethics}
In terms of ethical research conduct, we collected informed consent through the year 4 teachers from the students' parents or people with parental duties, which we accompanied with an extensive information sheet. We also included various options for withdrawing students from the research aspect of the study without missing the lessons (which formed part of the obligatory curriculum). All participants were free to withdraw from the study at any point on their own accord as well as by decision of their parents, guardians or people with parental duties. We received ethics approval for this study from Lancaster University's ethics committee (reference FASSLUMS-2023-2184-RECR-4). The researchers who delivered the curriculum further obtained a Disclosure and Barring Service (UK) criminal background check.

\subsubsection{Teacher Interviews}\label{ssec:interviews}
To ensure we could more conclusively reflect on the pedagogic and didactic elements of our delivery, we also chose to conduct semi-structured interviews with the two year 4 teachers subsequent to the delivery of the Ryelands AI Lab. The prepared questions ranged from general questions on AI and education as well as the experiences during the preparation, delivery and subsequent period of Ryelands AI Lab. The interviews were conducted by the first author and carried out first with each teacher individually for approximately 30 minutes, followed by a 15-minute group discussion so that questions emerging during the individual sessions could be addressed.\footnote{Question sheet and anonymized transcripts can be found at \url{https://osf.io/9afnr/?view_only=93c8d0e2159847dba2c454519682b85d}, accessed 24/04/2024.}

\subsubsection{Analysis}\label{ssec:analysis}
As mentioned above, we expected the exploratory study to produce ``intermediate-level knowledge,'' which could in turn be synthesized into ``strong concepts'' \cite{hook_strong_2012} that would inform the study's field; as well as more general implications from using RtD as a research methodology. In this, RtD's reflective capacity extended towards our patterns of analysis as well. As a general and oft-noted strength of design research---going back to Sch\"on's ``reflection-in-action'' \cite{schon_reflective_1983}---this meant that the weekly interplay between the initial curriculum, the occurrences of the preceding week, and the development for the upcoming week also shaped on-the-go analyses as well as a schema for findings. This can be further described, therefore, as a ``reflection-\textit{on}-action'' \cite{fitzgerald_theories_1994} beyond the immediacy of designing a particular thing. For instance, while we always assumed that the student's gain of particular practical skills would be an important measure of success for our approach, it only became clear through practice that a crucial qualifier for this measure would be the \textit{intentional} application of skills. We extended this activity-led analysis to the teacher interviews as well, which we considered from an ethnomethodologically informed perspective; meaning that their analysis was primarily directed by the first author's interpretation in relation to other considerations emerging from the delivery and design activities in this project. 

\subsection{Positionality}
In addition to the internal research ethics, we were mindful of further ethical dimensions of generative AI technologies, particularly with regards to the young age of students. While their capacities and the excitement they elicit are tempting, it is now widely understood that AI technologies can discriminate against particular protected categories (e.g., gender, ethnicity, nationality, sexual orientation), while also infringing on intellectual property by scraping content (e.g., data, images, texts), creating poorly paid and harmful jobs (e.g., content moderation, labelling), and contributing to the environmental impact of data centers and rare earth mineral extraction. As people without professional primary education expertise, this prompted the first two authors to reflect: is generative AI in schools a good idea in the first place, and are we---two white male Europeans---the people to introduce it to these students? Ultimately, we decided that for this set of students, the practical and critical competencies would aid them dealing with the increasing proliferation of their everyday lives by generative AI technologies. Further, we assumed that the intermediate knowledge we gathered would inform a more substantial critical position that we could add to the generative AI education space (see our proposition on the matter in section \ref{ssec:critical}) while foregrounding the benefits of constructionist approaches that combine practical and critical competencies.

\section{Exploratory Study: The Ryelands AI Lab}
This section details the actual instantiation of the Ryelands AI Lab, heuristically separated into its three significant phases. Again, it is important to note that these phases were only clearly identified in hindsight, and that elements within them overlapped substantially in practice.

\subsection{Pre-Deployment: Research and (Co-)Design}
First, we conducted \textit{initial research and co-design} activities with the year 4 teachers which produced the first lesson scripts and curriculum shape. This was followed by the \textit{design of learning artefacts} (i.e., image generation tool, slides, worksheets and secondary learning materials) both took shape from and in turn shaped the lesson scripts during actual delivery. Furthermore, all finalized design artefacts can be seen contextualized on a Miro board.\footnote{\url{https://miro.com/app/board/uXjVMw38td8=/?share_link_id=503687145119}, accessed 24/04/2024.}

\begin{figure}
  \centering
  \includegraphics[width=1\columnwidth]{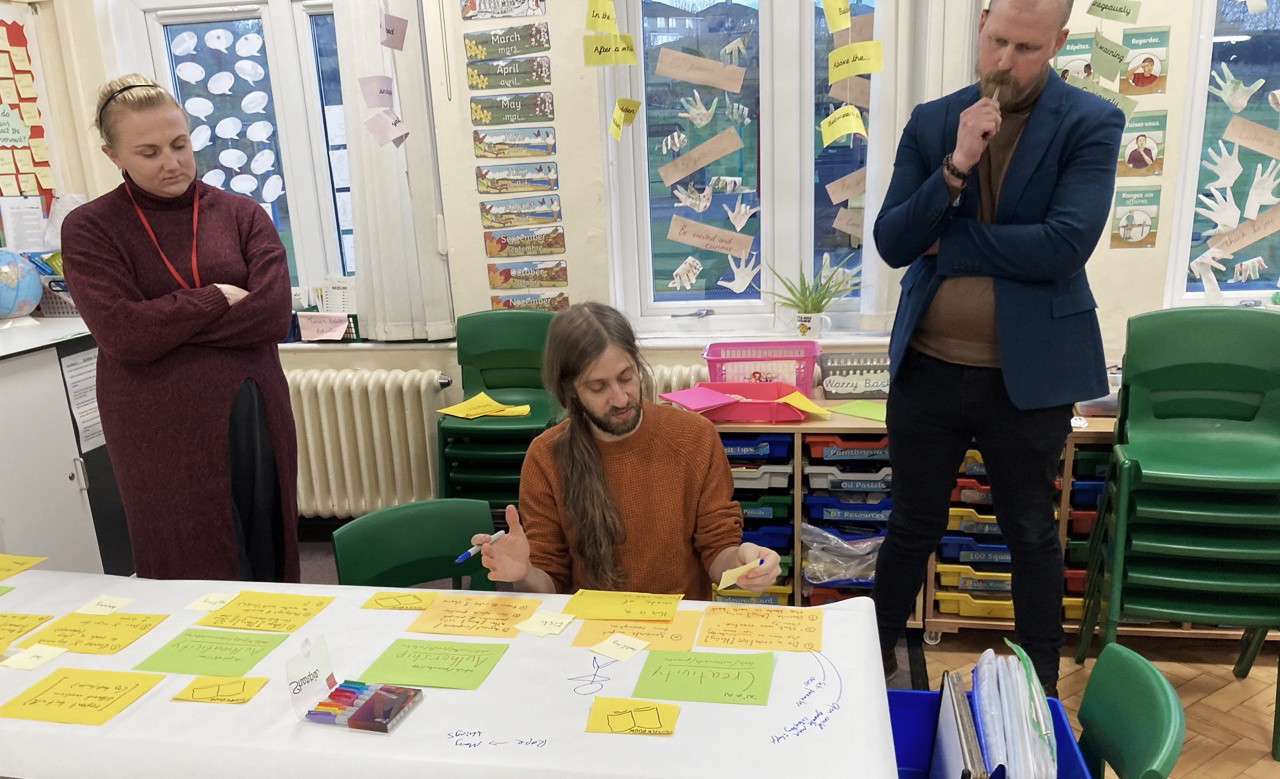}
  \caption{Image from early stage of co-design process, showing second author (middle) and the year 4 teachers.}
  \Description{Picture of three people in a classroom standing behind a table on which sticky notes, images, and sketches on curriculum ideas are spread over a long roll of paper. In front, second author is kneeling and in the process of explaining something on the long roll, to the left and right of them the year 4 teachers are standing and observing.}
  \label{fig:codesign}
\end{figure}

\subsubsection{Initial Co-Design}
From the start, we were mindful of supplementing the student's ongoing curriculum. To this end, we referred to the UK Department of Education's curriculum guidance; where we found that Keystage 2 Year 4 students' design and technology learning goals especially referred to the study of ``past and present design and technology [through which] they develop a critical understanding of its impact on daily life and the wider world''~\cite{noauthor_national_2013}. On this basis, we conducted four in-person and remote co-design sessions with the two year 4 teachers we contacted (see section \ref{ssec:approach}) over a period of three months. These sessions initially centered on demonstrations of what various generative AI technologies could do, before moving to mapping out important concepts and potential topics that would fit with the above learning goal (see Figure \ref{fig:codesign}). Given our overarching objective that students were to engage with an actual generative AI technology in order to develop practical and critical competencies, we identified multiple possible candidates for the latter---text-to-image (e.g., Dall-E), text-to-text (e.g., GPT3), image-to-text (e.g., CLIP) were all considered, as well as conversational interfaces (e.g., ChatGPT). We eventually settled on a text-to-image generation model as this seemed the most direct and literally generative way to engage students. The specific choice was Stable Diffusion \cite{rombach_high-resolution_2022}, an image diffusion model that was and continues to be highly prominent. Like other text-to-image generation models, Stable Diffusion is capable to use natural text (along with other parameters, see \ref{prototype}) as an input for a training data guided process of diffusing noise into an image reflecting the text input. Based on the image creation process taking center stage, topics to introduce in the lesson plan were mostly concerned with creativity and ownership, but towards the end also touched upon more complex issues such as generative AI technologies' transformation of reality (see \ref{ssec:delivery}).

\begin{table*}[t]
\centering
\caption{Overview of weekly delivery of lessons, with note on topics as well as materials and process.}
\label{tab:lessons}
\begin{tabular}{p{0.05\textwidth}p{0.15\textwidth}p{0.3\textwidth}p{0.4\textwidth}}
\toprule
\textbf{Week} & \textbf{Topic} & \textbf{Learning Outcomes and Approach} & \textbf{Notes on Materials and Process} \\ \midrule
1 & Introduction to AI image generation & It is possible to create photorealistic images based on text inputs, the chosen text impacts the resulting image. & Students filled in blanks in preexisting prompts with chosen words on worksheets, then were asked to enter these prompts into the image generation tool. \\ \midrule
2 & Refined prompt writing & Reinforcing basic prompt writing and learning how to use other parameters (e.g., diffusion seed, negative prompt) to help control image generation. & Students were asked to write their own prompts but were provided with some inspiration. They were also asked to experiment with advanced parameters and record their choices on worksheets.\\ \midrule
3 & Focusing on intention in prompt writing & Demonstrating and practising the use of descriptive language to translate specific intentions into prompts (e.g., a `blue door' vs `a door'). & Introduced the notion of a group project (`reimagining our school') as a means to focus prompt generation around a specific topic. \\ \midrule
4 & Advanced vocabulary & Practising how to use specialised vocabulary in complex prompts to achieve deliberate outcomes relating to a specific theme (`reimagining our school'). & The teachers did some `pre-teaching' prior to the classes where students were introduced to advanced vocabulary. \\ \midrule
5 & Ownership and styles & Showing particular styles of image can be obtained from AI image generators and that using artist names can accomplish this. Using this demonstration to explain how training data impacts upon generated images and opening a discussion on training data ethics and ownership of outcomes. & In this week the image generation tool was used to re-generate images from previous weeks using the same prompts, seed and other parameters but with the addition of keywords to achieve specific styles. Worksheets were used to capture students' thoughts on whether stylised images were preferable and on the ethical implications of creating images based on artist styles. \\ \midrule
6 & Retrospective, discussion, exploration of other types of AI & The final week was an opportunity to recap and assess previous weeks' learning outcomes. We also demonstrated other types of generative AI to assess whether the students could transfer their knowledge to other domains. & The final lesson mainly consisted of demonstrations at the front of the class. A worksheet was used to allow students to reflect on one of their chosen images, describing that image's attributes. \\ \bottomrule
\end{tabular}
\end{table*}

\begin{figure}[tbh]
\centering
\includegraphics[width=1\columnwidth]{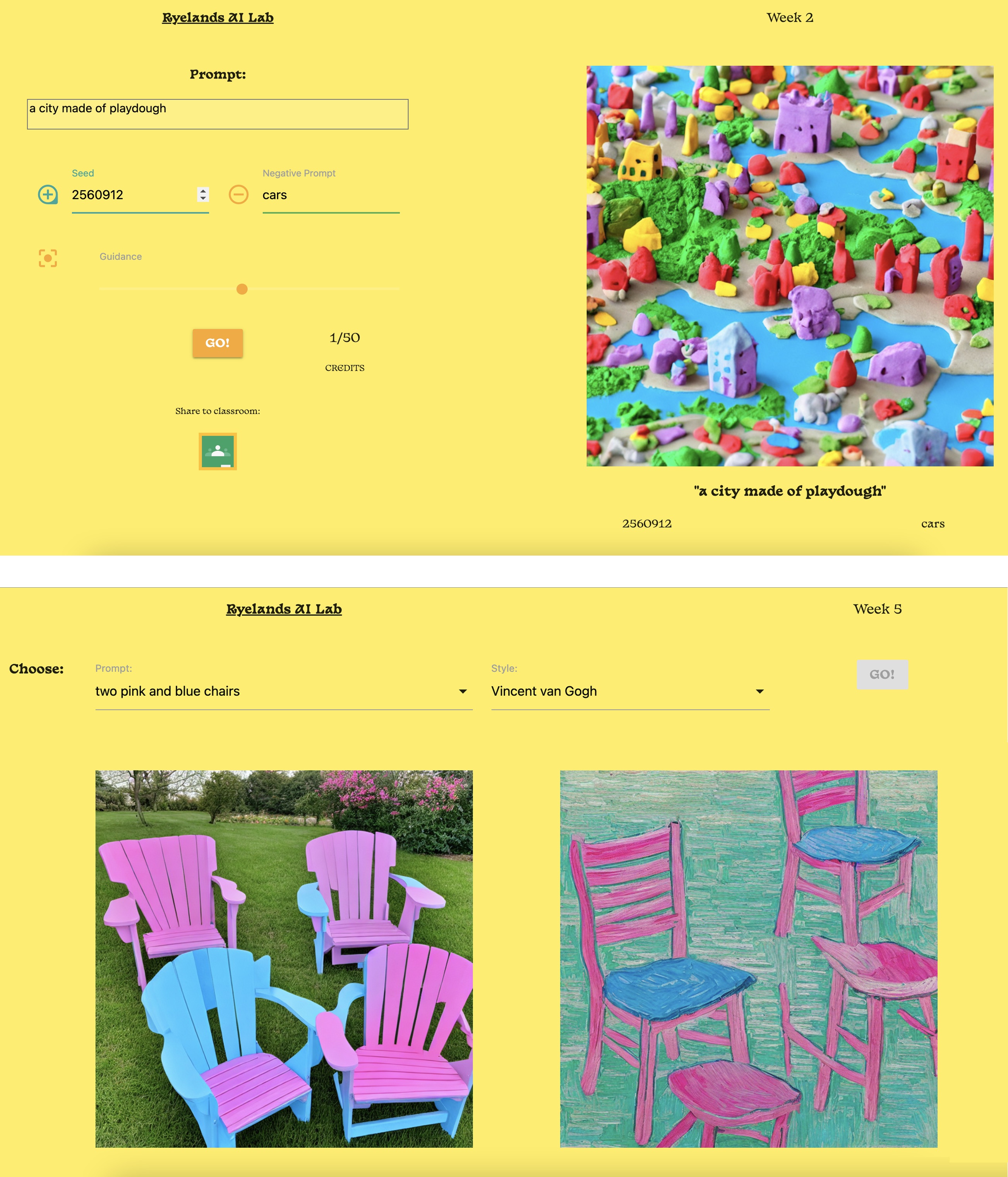}
\caption{Two versions of the image generation tool used during delivery. \textit{Top}: In this version used from week 2 onwards, students can input the main text prompt; and optionally specify negative prompt(s), an initial seed, and the context-free guidance scale, and submit to a custom Stable Diffusion API endpoint. \textit{Bottom}: A version created for reflection on issues such as ownership and creativity: students select from 5 randomly sourced preexisting image generations and their prompts and can `convert' these to a particular artist's style.} 
\Description{Two screenshots of the image generation web interface built for Ryelands AI Lab. Top: An interface with input options (e.g., textfields and sliders) on the left, and output (e.g., image, prompt, seed) on the right. In this version used from week 2 onwards, students can input the main text prompt; and optionally specify negative prompt(s), an initial seed, and the context-free guidance scale, and submit to a custom Stable Diffusion API endpoint. Bottom: An interface with input options in the upper third, specifically a dropdown on the left for pre-existing prompts and a dropdown on the right for artist styles. When the former is selected, the associated image to the prompt is displayed on the lower left; here, an image of 4 pink chairs on a lawn. Applying the artist's style, this image is converted to an output displayed on the lower right; here 4 pink chairs in Vincent van Gogh' abstract expressionist style.}
\label{fig:prototypes}
\end{figure}

\subsubsection{Image Generation Tool}\label{prototype}
We were aware that in class each student could make use of a Google Chromebook. This meant that we could design custom websites which afforded a less technical interface for accessing a generative AI technology than most current solutions, and allowed for direct interaction with the parameters of the generation process. The students used various versions of the image generation tool (HTML/CSS/JavaScript frontend, Jupyter Notebook backend) for interfacing with our chosen image generation model, Stable Diffusion via the platform replicate.ai.\footnote{\url{https://replicate.com}, accessed 11/08/2023.} In response to the lessons, there were four different interface design versions used over the six lessons (see Figure ~\ref{fig:prototypes} for two examples). In the first week, students used a simple version with textual input only, before moving on to a more advanced version that also included technical settings. In the latter, the seed (i.e., the initial noise image for diffusion), the negative prompt (i.e., textual input that discourages generation of particular things) and the context-free guidance (i.e., numerical value that steers the model away from or towards more randomness in relation to the prompt) could also be set. Starting from week 2, this advanced version was used by the students. In week 5, the website confronted students with an ethical question to be interactively explored: whether it was OK to use actual artists' `styles' for image generation. To this end, we adapted the website to randomly select 5 previous generations and let students choose an artists' `style' (e.g., van Gogh or Winifred Knights) for a direct comparison between original and `adapted.' To support interaction during delivery, we furthermore implemented a simple querying interface which allowed us to quickly display the image generations featuring a particular term and/or generation date. This way, we could gather students after a period of activity for discussion and reflection. Here, it is also noteworthy that an internal discussion on safeguarding students took place in the development of these web interfaces, led by the concern around explicit content that can be created with image diffusion models. To preempt this, we eventually decided to always frontload a student-initiated API request with a `hidden' negative prompt consisting of \textit{`gore NSFW creepy adult nudity horror erotic XXX weapon violence gun knife blood'}. This was a first practical indication of critical concerns around introducing generative AI technologies to minors, which we reflect on in our discussion (see section \ref{ssec:critical}).

\subsubsection{Further Learning Materials}\label{ssec:materials}
\begin{figure*}[t]
\centering
\includegraphics[width=1\textwidth]{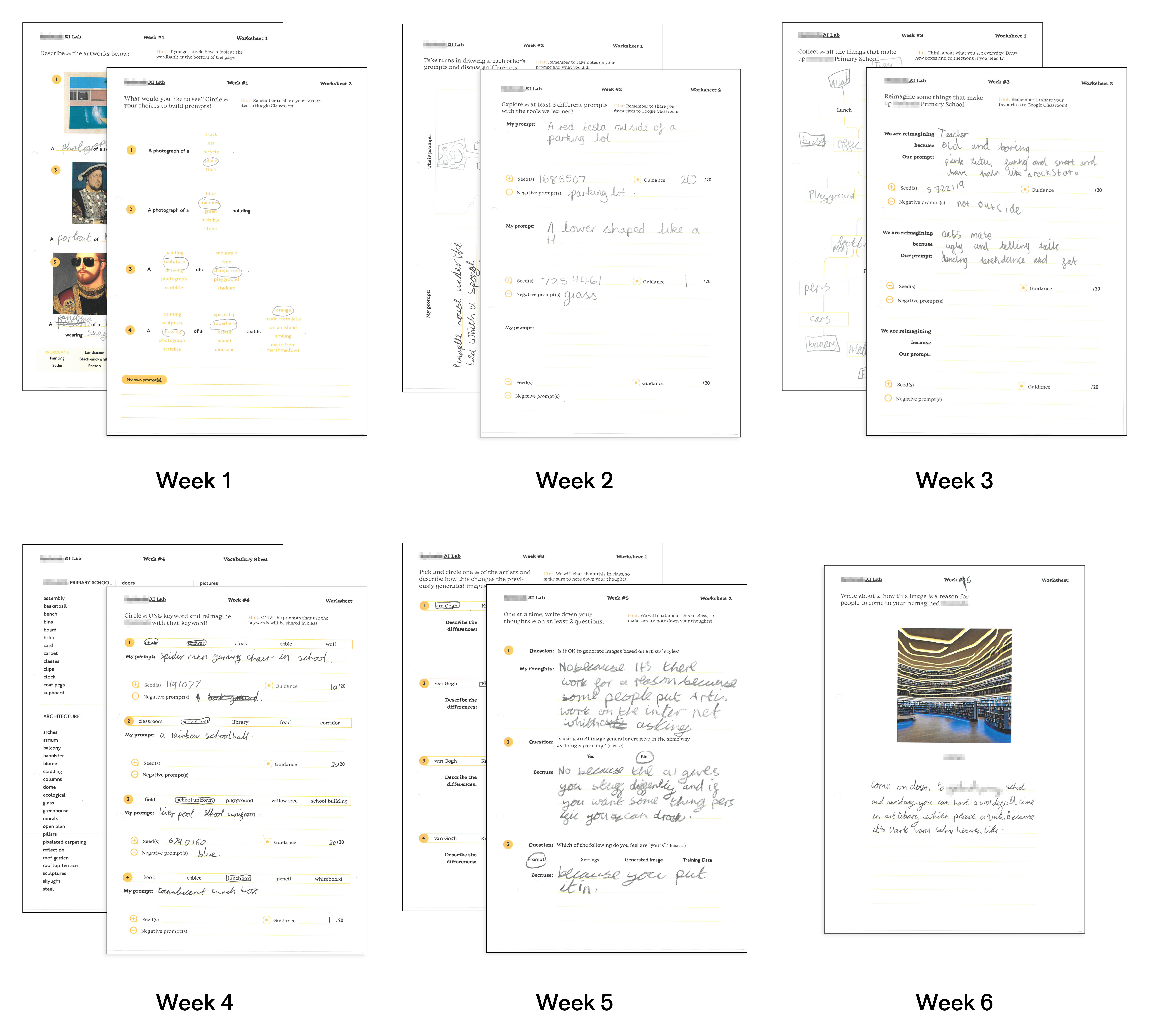}
\caption{Exemplary scans of the worksheets for each week which students would keep in folders. Weeks 1 and 2 feature an introductory worksheet first (e.g, ``Describe the artworks below'' in week 1). Weeks 3 and 4, with the introduction of the ``Reimagining Ryelands'' lab project, first inspires students (e.g., by prompting them to list all things they can think of relating to their school) and then asks students to log their reimagination experiments. Week 5 first asks students to log their experiments with artist' styles, and then provides a questionnaire for reflection. Week 6 features individual worksheets which show their chosen prospectus contribution and asks them to write a persuasive accompanying text.} 
\Description{Exemplary scans of the worksheets for each week which students would keep in folders. Weeks 1 and 2 feature an introductory worksheet first (e.g, ``Describe the artworks below'' in week 1). Weeks 3 and 4, with the introduction of the ``Reimagining Ryelands'' lab project, first inspires students (e.g., by prompting them to list all things they can think of relating to their school) and then asks students to log their reimagination experiments. Week 5 first asks students to log their experiments with artist' styles, and then provides a questionnaire for reflection. Week 6 features individual worksheets which show their chosen prospectus contribution and asks them to write a persuasive accompanying text.}
\label{folders}
\end{figure*}

Other learning materials took shape in weekly iterations in correspondence with the design of lessons and image generation tool. There were slides and, most prominently, folders for which students would receive weekly worksheets (see Figure ~\ref{folders}). The worksheets were generally employed to structure lessons by introducing students to basic concepts; for instance first describing an artwork, then using descriptions as a prompt to bring home the `back-to-front' logic of using an image description to create a previously nonexistent image. Additionally, the worksheets were designed to let students keep track of their generations and decisions by noting prompts, parameter settings, etc. Aesthetically, the image generation tool and learning materials developed together through the initial co-design phase, with choices of font and colour for instance constrained by browser affordances.

\subsection{Deployment: In-Person Delivery and Reflexive Adaptations}
Here, we detail the actual \textit{deployment} of our learning materials and scripts in the six lessons. 

\subsubsection{Lesson Overview}
While the overall thematic and conceptual pedagogical approach remained largely unchanged, actual lesson scripts, slides, worksheets and iterations of our image generation tool were designed, developed or adapted on a weekly turnaround basis. While this led to substantial time demands, it also allowed us to more directly adapt to actual needs and challenges which could not have exhaustively or even accurately been formulated prior to deployment. The final lesson plan is summarized in Table \ref{tab:lessons}. Additionally, it is noteworthy that the lesson foci shifted significantly following the introduction of the lab project `Reimagining Ryelands' in week 3, which we decided to pursue in response to the high quality of student engagement and to challenge students with a concrete goal relevant to their everyday lives. Subsequently, using the technical skills gained previously, the week 3-6 lessons gained a more constant and sustained focus.

\subsubsection{In-Class Delivery}\label{ssec:delivery}
\begin{figure}[tbh]
\centering
\includegraphics[width=1\columnwidth]{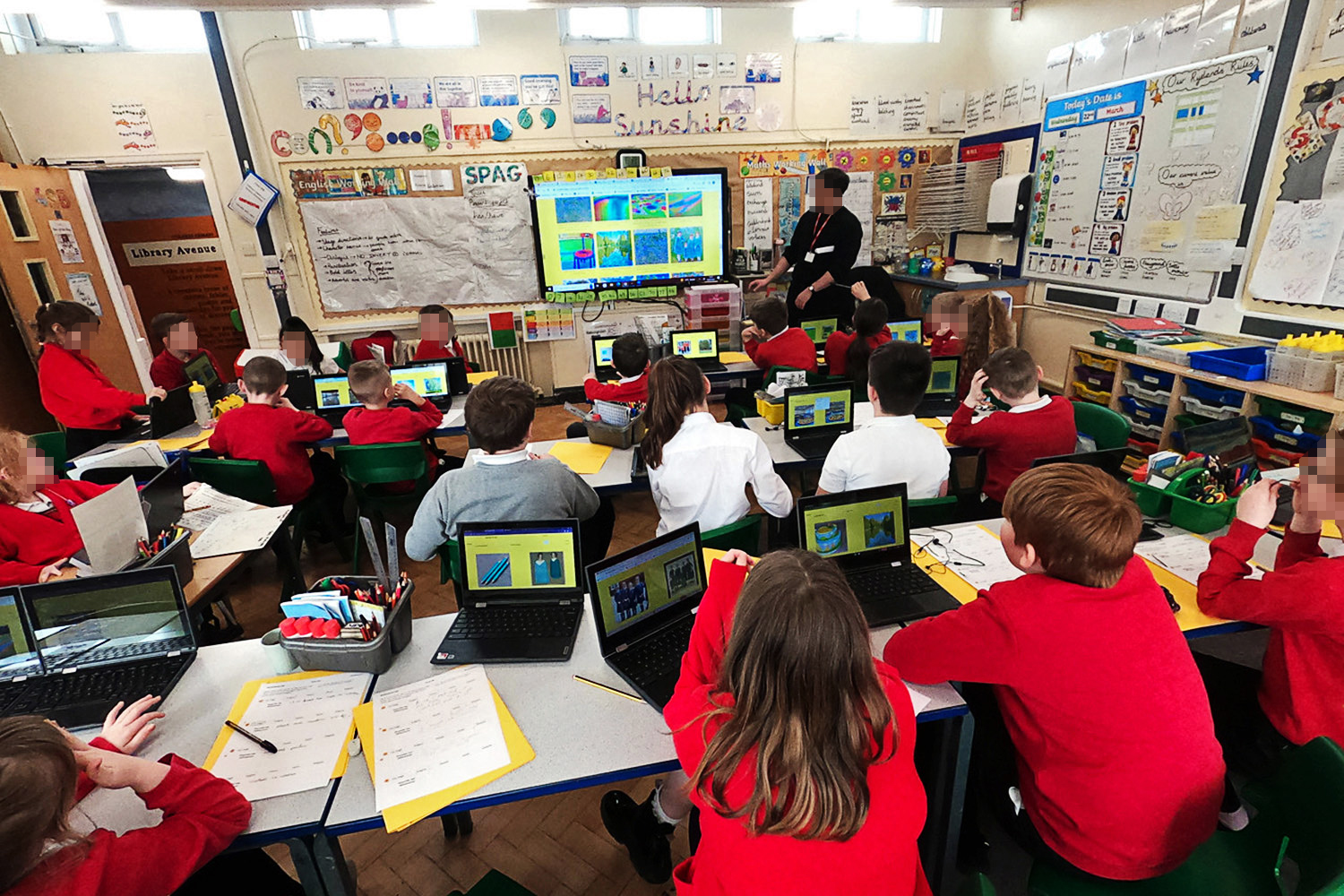}
\caption{Image from in-class delivery during discussion of students' creations in week 5, showing the `listing' interface on the whiteboard and the week's image generation tool (see Figure \ref{fig:prototypes}) on the students' Chromebooks.} 
\Description{Image taken from the back of the classroom from in-class delivery during discussion of students' creations in week 5, showing the `listing' interface on the whiteboard in the background with one of the authors pointing at it and the week's web interface (see Figure 3) on the students' Chromebooks in the foreground.}
\label{delivery}
\end{figure}

The two first authors delivered the six lessons along a timed but responsive lesson script according to each week, without specifying who spoke on what topic (see Figure \ref{delivery}). The modes of interaction generally proceeded from an interactive hands-up or call-out session, such as a game using the ``Which Face is Real?'' website,\footnote{\url{https://www.whichfaceisreal.com/}, accessed 23/07/2023.} to an instructional presentation period followed by a related worksheet. This typically more didactically focused half of the lesson would then be followed by more or less guided individual image generation tool usage on students' Chromebooks. The lessons were mostly concluded by reviewing recent image generations, which brought about the development of the `listing' interface for the tool (see section \ref{prototype}).

\begin{figure*}[t]
\centering
\includegraphics[width=1\textwidth]{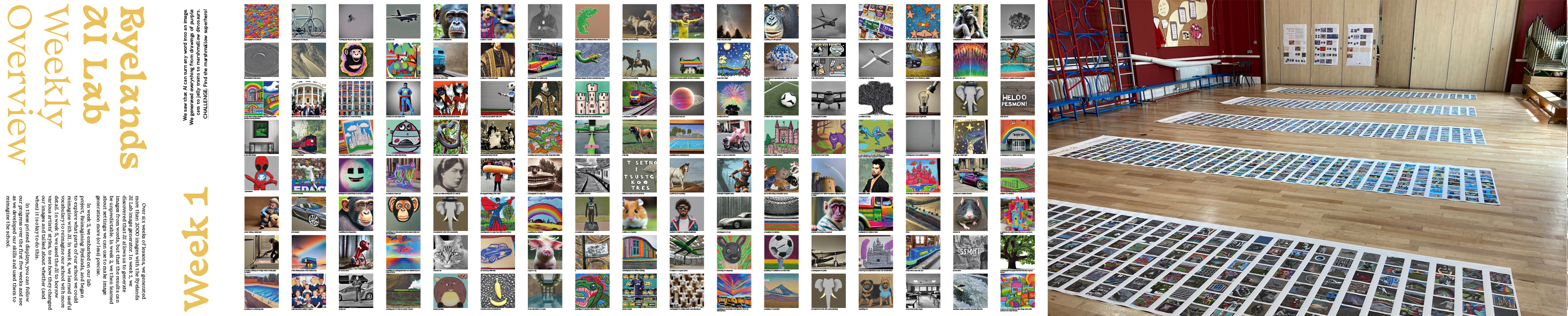}
\caption{Left: A cropped print file of the 5x1 meter strip showing an introduction to the week's delivery contents and a random selection of 250 images with their respective prompts. Right: Overview of the week strips laid out in the school gymnasium for the exhibition. In the background, three posters can be seen that show the spreads of the Ryelands AI Lab prospectus.} 
\Description{A composite of two images. Left: A cropped print file of the 5x1meter strip showing an introduction to the week's delivery contents on the left and a random selection of 250 images with their respective prompts in a grid arrangement on the right. Right: Photo of the week strips laid out in the school gymnasium for the exhibition. In the background, three posters are put up against a wall that show the spreads of the Ryelands AI Lab prospectus.}
\label{exhibition}
\end{figure*}

\begin{figure}
 \centering
 \includegraphics[width=1\columnwidth]{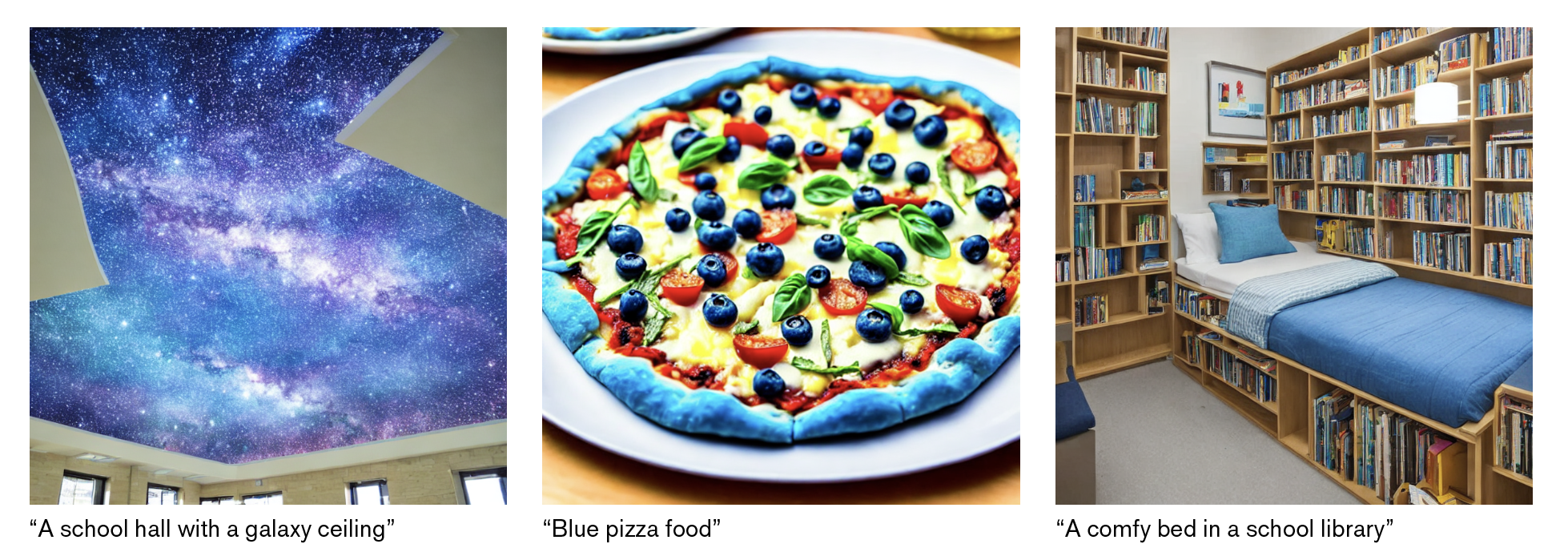}
 \caption{As part of the Ryelands AI Lab, students were given a lab project of ``Reimagining Ryelands''. The above shows three such reimaginations concerning their school hall, cafeteria food, and library using a generative AI text-to-image model, with respective prompts below.}
 \captionsetup{justification=centering,margin=2cm}
 \Description{As part of the six lessons delivered in the Ryelands AI Lab, students were given a lab project of ``Reimagining Ryelands''. The above images show three of such reimaginations. Left: A reimagined school hall showing a galaxy on the ceiling, prompt: ``A school hall with a galaxy ceiling.'' Center: Reimagined cafeteria food as blue, prompt: ``Blue pizza food.'' Right: A reimagined school library that integrates beds within book shelves, prompt: ``A comfy bed in a school library.''}
 \label{fig:labproject}
\end{figure}

\begin{figure}[tbh]
\centering
\includegraphics[width=1\columnwidth]{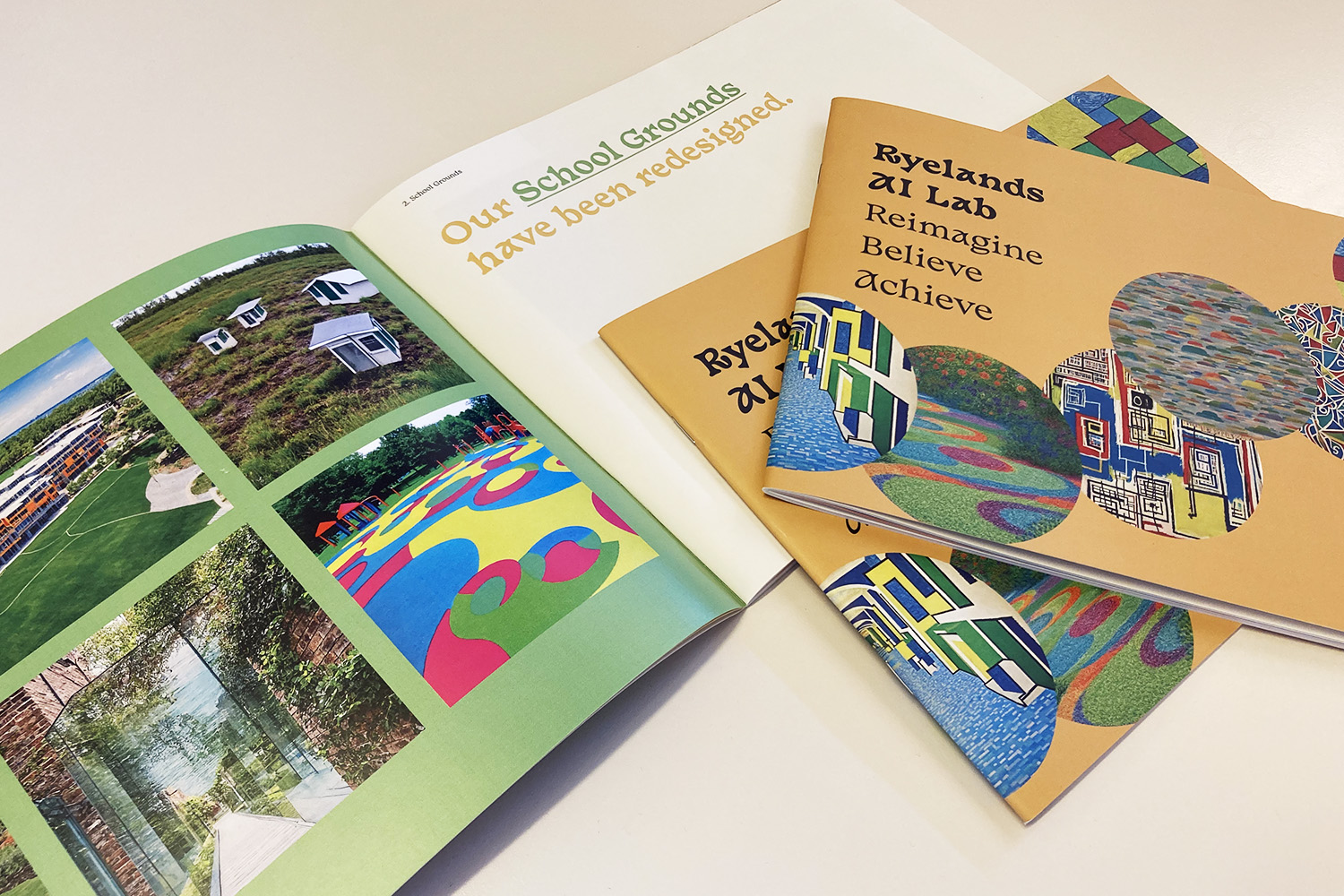}
\caption{The output of the ``Reimagining Ryelands'' lab project were transformed into an Ryelands AI Lab prospectus and printed copies were handed out at the exhibition.} 
\Description{An image of three copies of the Ryelands AI Lab prospectus, one open and showing a spread on the left and two closed and showing the cover on the right. The former has four student-generated images on the left relating to the theme of ``School Grounds'' on the left, and a section headline ``Our School Grounds have been redesigned.'' on the right page.}
\label{fig:prospectus}
\end{figure}

The year 4 teachers and teaching assistants remained present, and would support the delivery by, for instance, reformulating concepts in terms of previously encountered curriculum matter, or by focusing on a particular student according to emergent or ongoing needs. Particularly noteworthy is their activity outside of the Ryelands AI Lab lessons between week 3 and 4. Here, a custom vocabulary list with architectural and aesthetic terms was pre-taught to the students to support more specific and intentional generations. This reflected the student's architectural and aesthetic vocabulary deficiencies due to the Covid-19 pandemic, as well as their own desire for more `resources' for image generation. Further, in week 6 we delved deeper into examples that dealt with the effects of generative AI technologies, such as Bogost's article on wildfire skies turning gray in smartphone photographs \cite{bogost_your_2020}), and issues of representation (e.g., advanced `beauty' filters on social media \cite{bitmead_problem_2023}). 

\subsection{Post-Deployment: Processing and Sharing Outputs}\label{ssec:postdeploy}
In this section, we describe the activities undertaken after the delivery of the six-lesson-plan which consisted of processed outputs that we used for celebrating and sharing the students' achievements. It should be noted that the aim to hold an exhibition and to design an `imaginary' school prospectus only took shape during delivery in response to the outputs and ongoing questions.

\subsubsection{Exhibition}
The exhibition was designed to give a near-complete overview of the work students had put in. For each week, we printed 5x1 meter strips showing 256 randomly chosen images and their prompts for each of the five weeks in which image generation was center stage (see Figure \ref{exhibition}). We created a version of the image generation tool which incorporated all interactive elements (i.e., prompts, negative prompts, seeds, guidance-scale, artist styles) for the exhibition and set up interactive stations where students could show people with parental duties, siblings or other students their newly gained skills.

\subsubsection{Prospectus}
The design of the prospectus (see Figure \ref{fig:prospectus}) was inspired by existing materials found at other schools, such as info-brochures for parents/guardians as to what to expect from the school. In week 5, we asked students to pick one image to be included in this prospectus from among those created for the lab project `Reimagining Ryelands' from week 4 (see Figure \ref{fig:labproject}). In the following, final week, we further asked students to provide brief, persuasive descriptions of these images for the purpose of the prospectus. This material was then copy-edited into the Ryelands AI Lab prospectus, including an introduction by the two lead authors giving an overview of the project. The prospectus further included school group advertisements that integrated some of the common, unusual and amusing themes found in students' images (e.g., things made from marshmallows).

\begin{figure}[h]
  \centering
  \includegraphics[width=1\columnwidth]{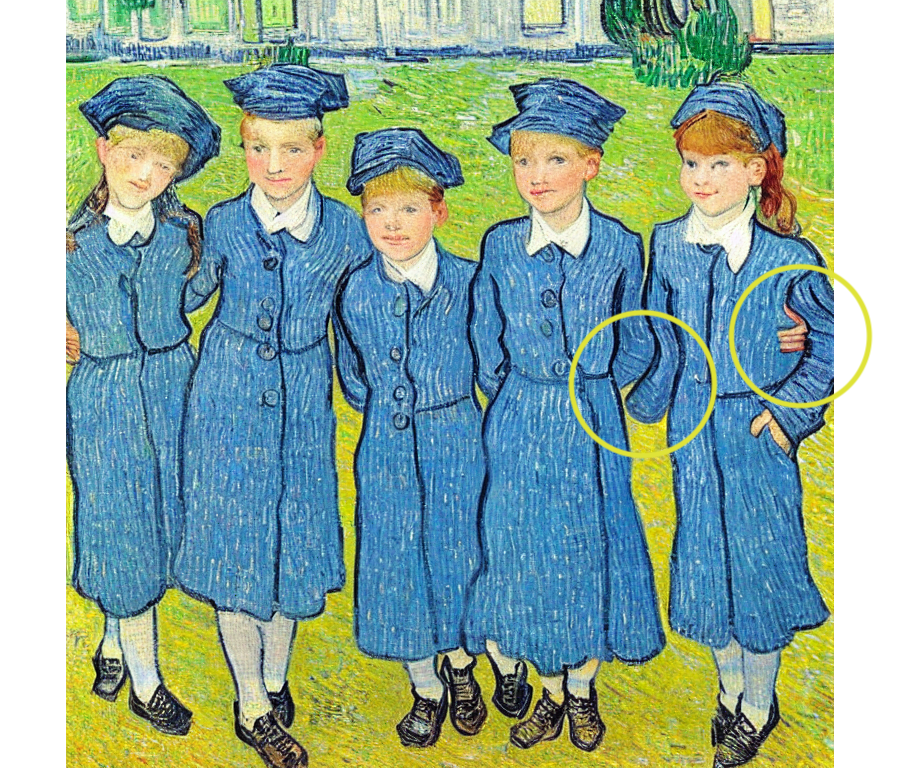}
  \caption{Example of an image brought up for discussion by a student in the week 5 lesson. Note the extra hand circled on the far right, and the empty sleeve circled to the left of the extra hand.}
  \Description{An AI generated image created in week 5 showing five school children in blue dresses, caps, stockings and black shoes in the style of Vincent van Gogh. The children are standing in line with their arms behind and around each others' backs, but on the right hand side a superfluous hand appears that has no relation to the arms in the image.}
  \label{fig:spottedhand}
\end{figure}

\begin{figure*}[]
\centering
\includegraphics[width=1\textwidth]{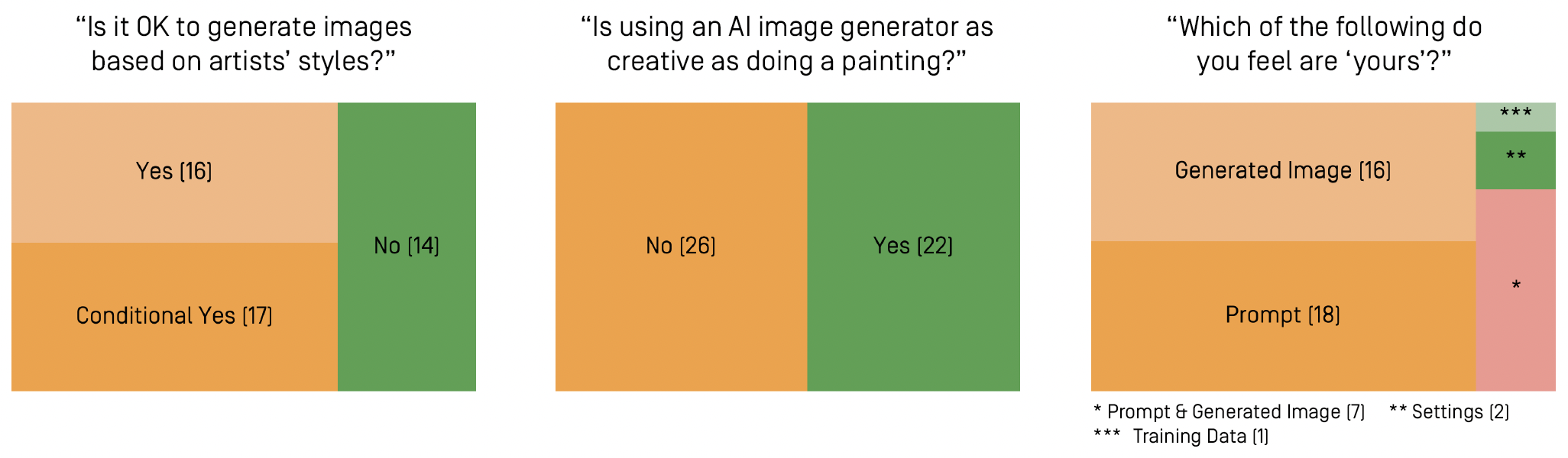}
\caption{Share distribution of student answers for the Week 5 questionnaire.} 
\Description{Treemap charts showing the share distribution of student answers for the Week 5 questionnaire. Left: Answers to the question ``Is it OK to generate images based on artists' styles?'': Conditional Yes (17), Yes (16), No (14). Middle: Answers to the question ``Is using an AI image generator as creative as doing a painting?'': No (26), Yes (22). Right: Answers to the question ``Which of the following (Prompt/Settings/Generated Image/Training Data) do you feel are `yours'?'': Prompt (18), Generated Image (16), Prompt and Generated Image (7), Settings (2), Training Data (1).}
\label{datagraphs}
\end{figure*}

\section{Study Findings and Generative AI Education Propositions}
In this section, we gather the findings generated within and through the exploratory Ryelands AI Lab study and close by synthesizing them in the form of ``strong concepts'' \cite{hook_strong_2012} (see section \ref{ssec:analysis}) that HCI can draw from as propositions related to the generative AI education context. Note that all findings are scaffolded by reflections of the teachers (T1, T2) gathered from the post-deployment semi-structured interviews, and that we have included descriptive elements to accurately represent the reflective interplay between a priori methodological decision-making and actual deployment of this project.

\subsection{Reviewing Constructionist Attributes of the Ryelands AI Lab}
Here, we consider a central constructionist aspect of our exploratory curriculum, and propose guidance on how HCI research in the educational context of generative AI technologies may build on our exploratory study.

\subsubsection{Findings on the Co-Development of Practical and Critical Competencies}\label{ssec:codevelop}
\begin{figure*}[tbh]
\centering
\includegraphics[width=0.8\textwidth]{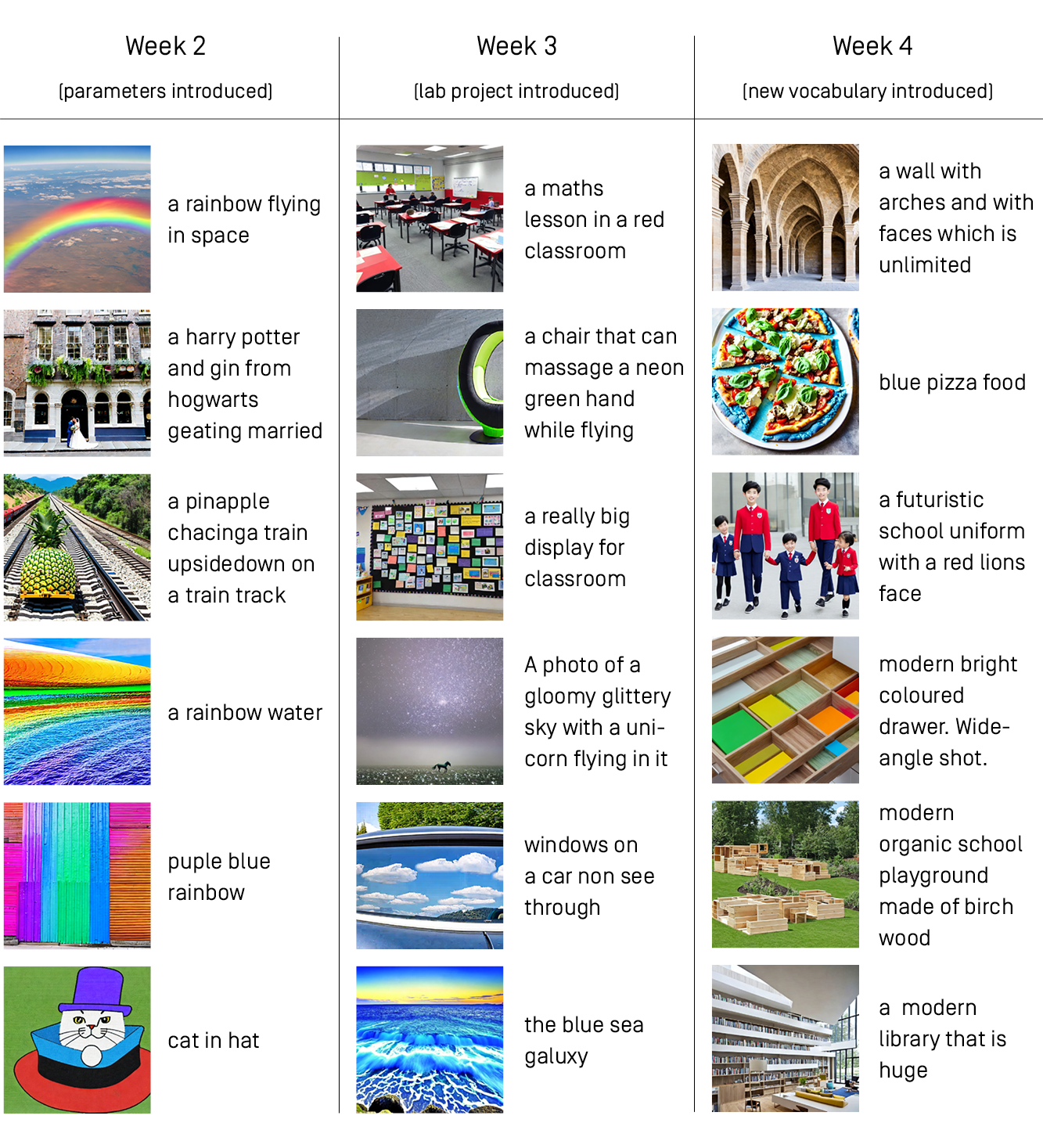}
\caption{Overview of exemplary, randomly selected images and their respective prompts from weeks 2 to 4. Note that even though the lab project was introduced in week 3, the lower three images show unrelated, more exploratory prompts.} 
\label{generatedImages}
\Description{A figure combining three columns of exemplary, randomly selected images and their respective prompts from weeks 2 to 4. The left column is headed ``Week 2 (parameters introduced)'' and features 6 images that show random contents, such as ``cat in hat'' or ``a rainbow flying in space.'' The middle column is headed ``Week 3 (lab project introduced)'' and features 3 images related to the lab project such as ``a really big display for classroom'' and 3 unrelated images such as ``the blue sea galaxy.'' The right column is headed ``Week 4 (new vocabulary introduced)'' and features 6 images related to the lab project with comparatively elaborated prompts such as ``modern organic school playground made of birch wood.''}
\end{figure*}

The major constructionist attribute of the curriculum is that we observed practical and critical competencies developing not distinctly, but rather exhibiting an interrelationship. Essentially, as practical know-how grew through continuous and diverse interactive sessions, students became more capable of reflecting on and even independently identifying critical aspects of their activity and the involved image diffusion technology. Over the first five weeks in the school the students created more than 2,000 images using corresponding prompts and parameters (week 6 was about reflection and discussion, so involved no image generation). To illustrate the diversity of the images the students created, Figure \ref{generatedImages} shows randomly selected images from weeks 2 to 4. Students' choices of prompts in the first two weeks seemed either random or unlikely things from everyday life or pop culture (e.g., Harry Potter, TikTok memes, footballers); providing a baseline for our observation of competency development. 

Following the introduction of the `Reimagining Ryelands' lab project in the third week, the image generations tended towards more deliberate choices. At the same time, this is not a clear-cut change as is shown by the three lower images in the week 3 column (see again Figure \ref{generatedImages}) which do not directly seem to relate to the goal of re-imagining the primary school. However, it can also not be ruled out that these were part of students' figuring out how to use the parameters (i.e., negative prompts, guidance scale, seeds) introduced in week 2, and thereby reflect the `model-probing' nature of constructionism---as students explored how far and where they can push a particular technology, they gain competency with the latter which in turn opens up more opportunities for intentional and reflective engagement. At this stage of the curriculum, the development of practical competencies (in terms of achieving desired, task-oriented results) indicated latent critical competencies in the probing interactions the former also supported.

In week 5, students used that week's image generation tool version (see \ref{fig:prototypes}, bottom) to re-generate previously generated images in particular artists' styles. This led to some students recognizing their previous prompts and noting the differences and distinct appearance of the artists' styles, which led to conversation among authors and students why that may be---particularly in contrast to some of their previous attempts to generate things familiar to them (e.g., their teacher but flying a plane). This allowed for a consideration of the celebrity of artists leading to many pixels of their work being on the internet and that these, unlike clearly identifiable portraits of \textit{their} teacher, were likely in the training data for the image diffusion model we were using. Most indicative is an example where a student excitedly asked to show their latest generation from this interaction on the class screen. The student showed the image in Figure \ref{fig:spottedhand}, where a superfluous hand appeared after the re-render in the style of Van Gogh. This led to a conversation in class on why the AI technology would show this hand, which, in combination with the previous conversation on artists' styles, led to discussing that AI technologies aggregate pixels rather than actually know what they generate. We argue that this is important to note: the student's growing practical competency with a particular image diffusion model made it possible to see the creation of images as more than `mindblowing,' but rather to scrutinize and question what appeared in front of them, which in turn opened opportunities for shared critical reflection on generative AI technologies. In other words, the familiarity and ease with which students then engaged the image generation tool made space for the strengthening of critical competencies such as observation and reflection.

After this engagement, we shared the story of artists' disgruntlement and anger with generative AI technologies.\footnote{Specifically, the case discussed here: \url{https://www.technologyreview.com/2022/09/16/1059598/this-artist-is-dominating-ai-generated-art-and-hes-not-happy-about-it/}, accessed 25/07/2023.} Then, we asked students (n=48 on the day) a series of questions on a worksheet, with two relating to ownership (``Is it OK to generate images based on artists' styles?'' and ``Which of the following [Prompt | Settings | Generated Image | Training Data] do you feel are `yours'?'') and one relating to creativity (``Is using an AI image generator creative in the same way as doing a painting?''). The results of these questions are visualized in Figure \ref{datagraphs}.

The responses for all three questions were ambiguous, but show consideration and reflection in this ambiguity. As is shown in the figure, a combined majority of students (n=33) stated that it was okay to use artists' styles, however of these a slight majority (n=17) further qualified this statement by saying it was only okay given a particular condition, such as the artist giving their consent or if they were dead. More idiosyncratic answers also argued that artists might like someone making pictures in their style. However, the students answering that it wasn't OK to use artists' styles were much less ambiguous when expanding on this selection, arguing that it would be a threat to their livelihood and theft. The results for the question whether generating images with an AI technology was as creative as doing a painting was equally ambiguous, with a slight majority (n=26) saying it wasn't. However, here again some answers were unexpected and not clear-cut, for instance some students said it was not as creative because they themselves had used the image generator, implying that they did not consider themselves creative. Concerning ownership, a slightly larger share of students (n=18) expressed feelings of ownership regarding the prompt, which ``came from my head.'' There was some overlap with feelings of ownership of the generated image (n=16), with some students (n=7) selecting both. The ambiguity (and indeed idiosyncrasies) in these responses indicate that students' competencies were not only repeated following instruction, but developed through the actual engagement with the technology at hand.

In the final week 6, we shared further controversial aspects of generative AI technologies as mentioned above (see section \ref{ssec:delivery}), but furthermore also acknowledged that we built in negative prompts to the image generation tool to keep the students, in our minds, safe from explicit imagery. This led to a different kind of conversation than we anticipated, with students and teachers reflecting on the similarities to search engines and the fact that, verbatim, if you search for upsetting things you will get upset; but also allowed us to point out the differences in that the image diffusion model could have generated upsetting things `unprompted.'

T2 describes the observed development towards a more reflective stance as a shift from ``the primary view [which is] basic systems stuff like `how do I make you change color and how do I make it bigger?' [towards getting] into the nuts and bolts of kind of like ownership and art and creativity'' (T2). They attribute this to the prolonged, hands-on approach of the project, stating that the students ``were able to find different layers to it with every single session, with every single session, their level of understanding of it grew'' (T2). 

In sum, these findings indicate (1) that practical and critical competencies of generative AI technologies cannot be clearly disentangled but, as per constructionism, co-develop; and (2) that the critical competency component of AI literacy, which may seem as the most challenging to convey to young people, can be significantly scaffolded by practical know-how. Building on this insight, we lay out our first proposition for HCI research to pursue constructionist generative AI education. 

\subsubsection{Proposition I: Constructionist Curricula for Co-Developing Practical and Critical Engagement with Generative AI Technologies}
We propose that the observed co-development of practical and critical competencies may encourage HCI researchers to utilise constructionist approaches to education in relation to generative AI technologies. Specifically, the instantiation of our exploratory study phases (see Figure \ref{fig:phases}) show that, as highlighted in other work, in-depth and prolonged hands-on engagement with generative AI technologies is needed to really pursue the development of competencies. For instance, the referenced work by Williams and colleagues notes that time and exploration leads to `better' performance \cite{williams_popbots_2019}, which are factors yet to be explored in the emerging approaches to generative AI education (e.g., the ``Prompt aloud!'' project by Lee and colleagues \cite{lee_prompt_2023} lasting for a single 90 minute lesson). Accordingly, we propose that the responsive study structure and our provided reasoning can assist HCI researchers and practitioners in various educational contexts (e.g., students in primary, secondary or tertiary education) that may on build constructionist-style education approaches such as ``experiential,'' ``active'' or ``inquiry-based learning'' (see \cite{kolb_experiential_2001,lau_review_1997,yannier_active_2021}, respectively).

First, during an \textit{initial research and (co-)design phase} prior to deployment, a shared baseline understanding with teachers can be established. The goal of this should not be to establish a concrete specification for what would happen next (e.g., fixed lesson plans or designs for the image generation tool) but rather to understand more deeply the underlying challenges relating to the unique elements of teaching generative AI competencies and how those challenges relate to the wider curriculum, pedagogic strategies used, and experiences with technology that students make in their everyday lives. It is noteworthy that while the teachers and ourselves were surprised by the quality and high degree of engagement of students during the second phase, this can also be considered more proactively to feed into the initial phases of future work. On reflection, we posit that despite their age, the students have grown up in a world dominated by the `reality-on-demand' logic of social media and streaming services. That a technology can serve up particular and strange images at whim is arguably a norm of this generation inasmuch as search engines---yet the particular differences to the latter should equally be stressed. It is therefore important for HCI research to more fully and precisely consider the competencies students can \textit{already} build on. Accordingly, future work may choose to intensify the participatory involvement of teachers in this phase (as Williamson argues, see \cite{williamson_degenerative_2023}) or by also involving students or people with parental duties to discern how `baseline' competencies relate to generative AI technologies.

Second, the actual \textit{in-person delivery and reflexive adaptations} of our exploratory study can be drawn from. As illustrated in Figure \ref{fig:phases}, the weekly lessons essentially formed responsive, interrelated components that fed into each other as the delivery progressed. We argue that the co-development of competencies was especially fostered in this regard. Facilitated by regular interactions directly with the image generation tool, the students seemed to have `small revelations' (e.g., if something appears in the image they don't like, negative prompts may remove it) which they would then experiment with further to find the limits of this new perspective (e.g., if the image generated from the seed just isn't a good one, then no amount of negative prompting will help). The example of the superfluous hand spotted and made into a topic of class discussion by a student above is a case in point: by that point in the curriculum, basic practical competencies were so far established that the creation of an image in the style of Van Gogh (in this case) was not the all-consuming focus, and therefore space for a further `small revelation' was made. In response, the authors could then adapt the content and form of the learning materials of the subsequent lesson accordingly. It is in this interplay that HCI researchers and practitioners can also identify where precisely to include more generalizable principles such as the relationship between training data, input prompts, and outputs that hold for generative AI technologies. If structured correctly, this has potential for providing the `sticky learning' that the teachers in our study referred to (see section \ref{ssec:teacherimpact}). 

We therefore suggest that responsivity and longitude are key attributes of a constructionist curriculum on generative AI technology, and that this may in turn actionably contribute to more expansive understandings of AI literacy---not as a single discrete set knowledge to be transferred (see \cite{long_what_2020}), but rather made of competencies to constantly develop and hone through prolonged, playful and `revelatory' engagement.

Third, the final phase of \textit{processing and sharing outputs} of the exploratory study was key to both explicate as well as celebrate the \textit{`flow of the work'} with our participants and related stakeholders; here students, siblings, and people with parental duties. To us, the significance of this phase lies in the framing of the students' achievements as core competencies \textit{for life}, rather than a one-off school art project with individual achievements. By giving students' work a concrete form (i.e., the imaginary prospectus built on the `Reimagining Ryelands' lab project) and a venue (i.e., the handing over of a prospectus copy for each student in an exhibition setting), another setting was provided in which students could reflect on and share the competencies they have built up; particularly with those in their lives who may take a longer view on how the developed competencies could help them navigate everyday life. This brings us to a final argument for the significance of HCI pursuing constructionist curricula on generative AI technologies. Seeing the latter as a contemporary ``foundational technology'' \cite{chameau_foundational_2014} that has already permeated much of the everyday activities of children (see \cite{pew_research_center_childrens_2020}), practical and critical competencies will be required of current and future generations as a prerequisite to twenty-first century citizenship. While we do not claim that this exploratory study has actually achieved this, we hope to have contributed to the emerging field that seeks to develop pedagogical approaches which consider the challenge of generative AI education in depth.

\subsection{Findings on Responsivity in the Curriculum Design Process}
Here, we stress the importance of responsivity in the design process of a constructionist curriculum on generative AI technologies to the particular educational context. This is shown by providing space to consider the teachers' perspectives on the Ryelands AI Lab, and their reactions to their own and their students' increasing competencies with generative AI technologies.  

\subsubsection{Perspective of Teachers}\label{ssec:teacherimpact}
The teachers reported that the Ryelands AI Lab project had a significant impact on them and their perspectives on teaching about technology on three levels. First, concerning their own knowledge of and attitudes towards AI technologies, they previously regarded AI in terms of how ``people are worried about it'' (T1) and as ``this kind of big, scary monster like skirting around the periphery of everything'' (T2). Early on, the teachers still felt that ``my mind was blown just by the very concept of what [generative AI technologies can do]'' (T2), whereas by the end T1 stated that ``I feel more confident to talk about [AI technologies], I can lead those conversations now'' (T1). Teachers delivering this type of curriculum intervention themselves is not as unlikely as may be thought. The realities of being a primary school teacher were evocatively described by T2: ``I am a 40-year-old man from Wigan. Is my French accent amazing? No. Will I teach French? Yes.'' (T2). This refers to the fact that especially primary school teachers need to gain confidence and expertise in subject matters which they do not have extensive knowledge of themselves. Both teachers referred to a whole industry that has sprung up to address this circumstance, mentioning ``existing models from, like, music companies [...] that literally give a whole package for teachers that are: `This is the prior knowledge you need to have''' (T2). The Ryelands AI Lab was seen as having potential to be adapted in a similar way, because ``the skeleton was able to be adapted so much and also the sharing of ideas and things enabled it to be really kind of a fluid approach and, and a really creative approach'' (T1). 

Second, from a pedagogical perspective, T2 highlighted the constructionist element of the curriculum, noting that it was ``really rare that we get to be in a situation where we know just as little as the kids do when we start off'' (T2). This led to them being ``able to [...] really lean into it with the kids, and their questions were our questions'' (T2). T2 further reflected that ``[this was] probably my first real experience sort of [this kind of mutual, constructionist learning] (T2). Concerning what it was that made this curriculum intervention successful, the teachers couldn't draw out one thing: ``It's like the demystification of the process, it's about the fact that it was a shared experience, it is about the fact that [...] you brought the kids in to be experts right from the beginning'' (T2). Akin to the description of the students' progress, they saw the intervention resulting in ``[having] the [actual technical] process of it demystified'' (T2); now thinking how ``AI is a bit of like an amalgamation of stuff that's already out there, which I didn't know before'' (T1). 

Third, the teachers began re-evaluating how AI and other technologies can be taught and to what benefit for their students. Concerning a more general media literacy encouraged by the curriculum, T1 stated that ``we've had a lot more in-depth conversations about art'' (T1) following the project. To them, the importance cannot be overstated, as ``we want them to look at things around them and not just take them at face value to actually question them because there is such a mental health issue with young people'' (T1), referring to social media and image filters specifically. This indicates that \textit{through} literacy gains surrounding AI technologies, other aspects of media (in this case, images) can become targets for more reflective engagement as well.

In sum, the impact on the teachers reported here is indicative that a constructionist, hands-on engagement with generative AI technologies is not only valuable within its own lifetime and solely for students, but also seemed valuable to the teachers, helping them to reflect on their teaching strategies. While the above quotes are essentially exclusively positive with regards to the developed curriculum, we use them as an impulse for our proposition that . Further below, parts of this finding also inform limitations we see for future work (see section \ref{ssec:limitations}).

\subsubsection{Proposition II: Approaching Generative AI Technology Education with Critical Responsivity}\label{ssec:critical}
Above, we documented how the teachers framed the responsivity of the curriculum `skeleton' and the gradually more complex hands-on engagement as an overwhelmingly positive experience. However, from a more distanced perspective and regarding our experience as researchers, we here reflect on the opportunity for more critical engagement in design that we and other HCI researchers may need to pursue. As mentioned above, generative AI technologies in educational contexts are frequently seen through either a punitive (e.g., preventing students from cheating) or utilitarian (e.g., improving teacher efficiency) lens. Our practical experience with the Ryelands AI Lab has brought further concerns to the fore, which connect the generative AI education space to more critical scholarship such as calls for a ``decolonial approach to AI in higher education teaching and learning'' \cite{zembylas_decolonial_2023}. In short, aside from the above argument for intertwining the development of practical and critical competencies \textit{within} a constructionist curriculum, we propose that the design of the latter calls for `critical responsivity.' By this we mean that researchers engage in an ongoing noticing of the emerging concerns of stakeholders which crucially could not have preceded the actual interaction with a generative AI technology. 

The difficulty in critically negotiating the capacities of generative AI technologies in an educational context came to the fore for us swiftly in the form of our `frontloading' of negative prompts (\textit{`gore NSFW creepy adult nudity horror erotic XXX weapon violence gun knife blood'}). While this `need' was shaped by our particular participants, it is noteworthy that it was \textit{our} need---whereas the teachers were at ease with considering it similar to the use of search engines by students. At the same time, this could have provided us with an opportunity to stress in which ways this ease may need to be tempered, given the difference between targeted search and the barely noticeable ``pattern leakage'' \cite{benjamin_machine_2021} that may crop up in AI technologies due to their probabilistic abstracting from training data. That we did not act on this opportunity is, in hindsight, an indicator where `critical responsivity' could have informed our design of subsequent lesson materials to act on this faultline. Similarly, a limitation of the current set of learning materials (particularly, the more instructive slides) is that more `sharp end' examples are needed to represent and educate on harms brought about by AI technologies not dissimilar from the kind of technology the students were using. While there is potential for more in-depth discussions to arise in a possible multi-year programme (e.g., a year 4 module later supplemented by an advanced year 6 module) as advocated by the teachers,\footnote{This would also allow for an introduction of more intricate and challenging aspects of AI technologies and society, such as shown in the works of, for example, Bender and colleagues \cite{bender_dangers_2021} or Birhane and colleagues \cite{birhane_multimodal_2021}. Such work would help to further underline the economic motivations behind much of AI technological development (e.g., the human element of training data production, extraction, curation and/or moderation), and also distinguish educational from utilitarian efforts aimed at `efficiency' such as learning how to write prompts in order to capitalize on generated images.} there were multiple opportunities for a more attuned ad hoc response. For instance, this could have been to stress that we were using an `off the shelf' image diffusion model (i.e., non-finetuned Stable Diffusion); and discuss with students in which contexts such generic solutions would not work---then extending it considering harms of generic models in arguably higher stakes contexts such as facial recognition, criminal recidivism prediction or credit scoring. This echoes concerns from critical voices such as Nemorin and colleagues who have noted that often little attention is paid to how AI technologies in educational contexts may reify knowledge along extractivist lines \cite{nemorin_ai_2023}. 

In sum, the ambition for ``lifelong learning'' \cite{eynon_methodology_2021} regarding AI technologies needs to bring values beyond the purely economic to the forefront. We argue that in the long term, constructionist HCI research efforts may help in addressing this need, but the intricate nature of critical concerns and rapid changes also necessitate action in current design efforts. In practice, and reflecting on our disclosure of the frontloaded negative prompt and the subsequent exchange with students, we suggest that HCI research could be proactive and responsive here by adopting a practice of \textit{disclosure}, rather than optimization. A parallel can be drawn here to the field of ``explainable AI'' (XAI), particularly regarding how Benjamin and colleagues' have stressed that while machine learning uncertainty is usually explained or engineered away, it can serve as a design material in its own right when being actively engaged \cite{benjamin_machine_2021}. In other words: in addition to putting `safety' of stakeholders first (e.g., by adding locked negative prompts to an image diffusion model), the disclosure of having done so (i.e., of having \textit{further} biased a biased model) can function as a way to critically ground reflection on the capacities and dangers of generative AI technologies. The challenge and potential remedy, we propose, in pursuing this lies in researchers---or educators---designing generative AI technology education provisions with critical responsivity in mind.

\section{Discussion}
In this section, we present a discussion relating to broader implications for HCI researchers regarding firstly the design of generative AI systems, and secondly the methodological strength of RtD to respond swiftly to disruptive technologies.

\subsection{Constructionist Strategies in the Design of Generative AI Systems}\label{ssec:constdesign}
While the Ryelands AI Lab carries direct implications for constructionist AI education, there is also intermediate knowledge we can build upon in this regard for more general design concerns. Here, we consider two primary aspects of designing generative AI systems where constructionist strategies may support HCI: first, during development as a way to build understanding and requirements of a system; and second, as guidance offered in the design of artefacts (e.g., onboarding user interfaces). Both aspects connect to the larger discourse on XAI or `interpretable machine learning (ML),' but importantly weigh the importance of explanation and interpretation differently. Similarly to the difference of constructionist to `instructionist' (e.g., frontal lecturing, passive students) learning methods, the goal of designing systems using constructionist strategies does not lie in transferring one definitive explanation or interpretation but rather in providing opportunities for engagement; whether this be playful or serious. To clarify the below, we consider constructionist strategies as ways to foreground particular \textit{affordances} of a technological artefact; i.e. the relations through which the world and the ways in which to perceive and act in it takes shape \cite{norman_affordance_1999, ingold_back_2018}.

First, constructionist strategies may benefit both designers and stakeholders in the early phases of system development, such as in  participatory co-design methods or place-based inquiries. Already in and of themselves, these types of methods are not generally present in XAI research, which predominantly engages with expert audiences, specifically people with formal ML education (e.g.,~\cite{liao_questioning_2020,madaio_co-designing_2020}). Following  proposals for the field to shift focus (cf.~\cite{wang_designing_2019,ehsan_human-centered_2020}), work in XAI has begun to diversify its methods and audiences. For instance, Benjamin and colleagues have conducted co-design workshops with employees at a research institution where the goal was not to convince stakeholders of the suitability of a proposed ML-driven visualization system, but rather to understand how precisely explanations for that system's outputs shape the highly specific contextual understanding of the stakeholders, with participants literally constructing representations of their context from materials such as playdough, which in turn led their analysis for the design of their system (see \cite{benjamin_explanation_2022}). Such constructionist strategies therefore further stand in contrast to the generally `instructionist' stance of XAI even when non-experts are considered, where stakeholders are often left unclear about how generic explanation metrics such as fairness map to their specific context (see \cite{saha_measuring_2020}). In this vain, we argue that constructionist strategies which closely couple opportunities for practical and critical engagement may further assist HCI in developing and pursuing such context-aware methods.

Second, constructionist strategies can also serve in the design of system \textit{artefacts} such as onboarding interfaces that can help stakeholders comprehend the particular generative AI technology's capacities rather than disguising or obfuscating them. Such HCI work is becoming more pressing, for instance to contextualize and weigh the outputs of text synthesis models such as ChatGPT that are particularly prone to ``fabrications and falsifications'' \cite{emsley_chatgpt_2023}. However, it is unlikely that there will ever be a perfectly trustworthy form of presenting current generative AI technology outputs---when developers themselves can frequently not fully account for their systems and/or routinely misjudge and over-accept explanations for AI technologies \cite{springer_dice_2017,kaur_interpreting_2020}. Thus, we suggest that constructionist strategies for generative AI systems offers a plausible route forward: rather relying on explanatory elements which puts people interacting with a generative AI system into a passive role, being given the means (e.g., through interactive examples) to playfully work through generative AI technologies' capacities within a given system may empower stakeholders. For instance, in contrast to the highly specialized explanatory interfaces dominant in the field (cf. ~\cite{hohman_gamut:_2019}), a constructionist design strategy may pursue incremental increases in the complexity of an interface (e.g., by introducing parameters such as seeds or guidance along specified routes). This could firstly support the practical understanding of a generative AI technologies' capacities, and secondly scaffold the multitude of ``explanation styles'' ~\cite{binns_its_2018} that people bring with them by allowing various thresholds for reflection. In this light, we argue that design can learn from educational approaches more generally to design for playful, `tinkering' engagements, and note that importantly such `ludic' design of generative AI systems can build on a long research tradition tracing back to Gaver and colleagues' \textit{Drift Table} \cite{gaver_drift_2004} and contexts such as playful encounters with robots \cite{lee_ludic-hri_2020}.

\subsection{Research-through-Design as a ``Rapid Response Methodology''}\label{ssec:RtDRRM}
Our main methodological interest was the effects of choosing RtD as our guiding project methodology for engaging generative AI technologies in this context. The relative openness of the topic of generative AI technologies in terms of its novelty and connections to all kinds of subject areas in the UK curriculum (e.g., art, citizenship, design, engineering) led to a burgeoning of design possibilities. During the initial collaborative process, this led us to envision a whole range of generative AI technologies (particularly text models such as ChatGPT) to be included in the delivery. Up until the first delivery, this was seen as feasible and the first lesson plans as well as scripts included details on how the various technologies would tie into each other. For instance, in one version the students were to create an avatar of some sort (e.g., a superhero) using a text-to-image model, then write scenes for the character using a text prediction model, which then would be used for further image generation. Additionally, we also considered making a hardware component; specifically a micro:bit\footnote{\url{https://microbit.org/}, accessed 01/08/2023.}-based camera which would leverage Stable Diffusion's image-to-image mode for an even more direct (and less desk-based) hands-on engagement with the AI technology based on Benjamin and colleagues' \textit{Entoptic Field Camera} \cite{benjamin_entoptic_2023}---with us thinking that these \textit{artefacts} would be the focus of the RtD methodology. As stated repeatedly above, these initial plans did not `survive first contact' with the realities of introducing fundamental concepts alongside the `mind-blowing' effect of the sheer possibility to create never-seen-before images from words. While we, therefore, did not achieve the same scope of topics and technologies we had initially sought out, we could rapidly adapt to the flow of the interventions rather than having to stick to a pre-given and potentially less impactful plan. Further, the RtD methodology we followed allowed for a previously unplanned introduction of a lab project (`Reimagining Ryelands') and the subsequent creation of the prospectus and exhibition.

This experience now leads us to consider the choice of RtD as our project methodology. The `inward' responsivity of RtD projects and related design research methodologies is well known, and has been extensively detailed above in terms of weekly adaptations of learning materials, reshaped project goals, etc. Here, we reflect briefly on the \textit{outward} responsivity of RtD. To put this into perspective: the text-to-image diffusion model we used was released only months prior (August 2022\footnote{\url{https://stability.ai/news/stable-diffusion-announcement}, accessed 24/11/2023.}) to our first meeting with teachers (November 2022). We find this noteworthy due to the vast space of concerns that was opened up in short succession to the release of it and similar models (e.g., ChatGPT, Midjourney, etc.)---as highlighted in the Writers' Guild of America strikes \cite{noauthor_wga_2023}, or a UK House of Commons committee report \cite{culture_media_and_sport_committee_connected_2023}. We argue that it was exactly this rapidly unfolding and ongoing event which made it feasible for us to choose RtD as a methodology. That is not to say that the widespread integration and proliferation of generative AI technologies won't require large scale studies---however, given the limitless contexts in which generative AI technologies as \textit{software} artefacts can be applied, HCI also needs to be able to quickly set up, deliver, and understand small-scale and single context studies. 

We argue that RtD is eminently suited to study the impacts of disruptive technologies while their socio-cultural-technical ramifications are still being negotiated. This is likely particularly true with AI technologies, where for instance the gap between user-facing apparent `intelligence' and vast socio-technical realities (e.g., content moderation, bias, data sourcing) can be particularly pronounced. At the same time, any technological innovation or adaption brings about uncertainty regarding its socio-cultural ramifications. This fits into a strong tradition within RtD and HCI design research where ambiguity \cite{gaver_ambiguity_2003} or multiple meanings \cite{sengers_staying_2006} are seen as resources rather than obstacles. Yet, we also think that the \textit{outward} responsivity of RtD is perhaps obvious, but underspecified. In turn, we propose to term this attribute of RtD as its suitability for being a \textit{rapid response methodology}, a term we borrow from museum studies. ``Rapid response collecting'' was initiated in 2013 as a curatorial strategy by The Victoria and Albert Museum (see \cite{etherington_interview_2013,victoria_and_albert_museum_rapid_2017}) to deal with significant quickly unfolding events in an increasingly connected world saturated by scaling information technologies---for instance, 3D-printed guns.\footnote{More recently, rapid response collecting resurfaced during Covid-19 pandemic focusing on the related paraphernalia and side-effect of widespread shifts to online socio-cultural activities \cite{debono_collecting_2021}.} In contrast to rapid response collecting, where museological ordering and contextualizing takes precedence, it is the strength of RtD to synthesize intermediate knowledge from the materials it gathers that is not only reflective but \textit{generative} towards new theories, products, design strategies, or subsequent research. But in contrast to other types of design processes (e.g., the double-diamond framework), RtD here can also be seen as responding to the questions thrown up by \textit{historical} phenomena rather than the demands of a product for particular ends (e.g., efficiency, UX, profit). This becomes all the more pressing given the general difficulty in preventing negative outcomes of technology once these have been identified (often termed the `Collingridge dilemma,' see \cite{kudina_ethics_2019}), and specifically the already significant outpacing of regulatory efforts by releases of AI technologies (see \cite{satariano_how_2023}). Accordingly, we argue that seeing RtD as a rapid response methodology can lend further legitimacy to the studies and researchers subscribing to it while still maintaining its specificity as a research methodology first and foremost.

\subsection{Limitations and Future Work}\label{ssec:limitations}
One of the main limitations of this project is its relatively small sample---one exploratory study in one school with one specific set of teachers and students. As such, it cannot be ruled out that parts of our observations came about due to the ``novelty effect'' (see \cite{agarwal_time_2000,mirnig_blinded_2020}) of generative AI technologies for students and teachers. While we think that it is exactly this novelty which needs to be addressed, and that RtD is in a position to do so (see section \ref{ssec:RtDRRM} above), it remains a concern within the context of generative AI education approaches. This effect may be alleviated with further studies, for instance conducting a more complex intervention with the same year group of students at a later date to assess the sustainability of our approach; or with studies that are explicitly designed for triangulation to test the approach for scalability. Regarding the latter specifically, and with encouragement from the teachers, we are planning to extend the approach to other schools in the area and other age groups (e.g., secondary and high schools). Given that one of the main learning artefacts, the image generation tool, can be accessed very easily by any device capable of web browsing, such efforts could focus on \textit{scale} primarily---there are as many as 16,783 primary schools in England alone.\footnote{\url{https://explore-education-statistics.service.gov.uk/data-tables/permalink/74bb0ee9-712c-4820-ad0e-08dbb395de42}, accessed 14/09/2023.} However, there are multiple design challenges such future work needs to navigate, especially given that it would mean pivoting from our responsive approach towards building the kind of package which are provided by platforms to rapidly upskill teachers on particular subjects, such as Twinkl.\footnote{\url{https://www.twinkl.co.uk/}, accessed 03/08/2023.} To frame this challenge productively, we are considering to frame it through the established HCI lens of ``research products'' \cite{odom_research_2016} developed by Odom and colleagues and recently expanded precisely in the dimension of ``scale'' \cite{boucher_research_2023} by Boucher, as this clarifies the particular demands of scaling the Ryelands AI Lab by articulating a design space with constraints. This would also allow for further exploration of a specific RtD concept as a way to rapidly respond to emergent \textit{design} challenges of disruptive technologies. However, a packaged module could also easily become just another generic resource commodity pushed on overworked and underpaid educators without enough critical perspectives or representative examples; which would have to be carefully considered and critically reflected upon (see also \cite{gulson_repackaging_2022}). 

\section{Conclusion}
In this paper we detailed how we used a Research-through-Design project methodology to engage the uncertainties surrounding generative AI technologies with an exploratory study in the educational context; which took the shape of the Ryelands AI Lab as a constructionist pilot curriculum for generative AI primary education. Adopting RtD's capacity for inquisitive and reflexive development, we designed, produced, and delivered six lessons on generative AI technologies which appears to demonstrate improvements in practical and critical competencies for students; and drew intermediate knowledge from our observations for future work in HCI. Concerning the project's context directly, we (1) offer guidance for HCI research on developing constructionist generative AI curricula, and (2) critically reflect on the role of AI technologies in education based on our experience. Reflecting on wider implications for the field, we further (3) consider the value of constructionist strategies in designing generative AI systems and (4) look at RtD as a `rapid response methodology' that is particularly suited to unfolding and unsettled socio-technical developments surrounding disruptive technologies. Especially this latter aspect shows, we argue, why design can attend to emerging uncertainties in unique and valuable ways.

\begin{acks}
We cannot thank the amazing children of classes 4SB and 4G enough for their enthusiasm, curiosity and wild imaginations. We also want to thank headteacher Mrs Linda Pye for her support and encouragement in making this project happen, as well as Willow Mitchell and Joe Bourne for their invaluable input. This work is supported by UK Research and Innovation (grant MR/T019220/1, ``Design Research Works'').
\end{acks}

\bibliographystyle{ACM-Reference-Format}
\bibliography{bibliography}


\begin{thebibliography}{99}


\ifx \showCODEN    \undefined \def \showCODEN     #1{\unskip}     \fi
\ifx \showDOI      \undefined \def \showDOI       #1{#1}\fi
\ifx \showISBNx    \undefined \def \showISBNx     #1{\unskip}     \fi
\ifx \showISBNxiii \undefined \def \showISBNxiii  #1{\unskip}     \fi
\ifx \showISSN     \undefined \def \showISSN      #1{\unskip}     \fi
\ifx \showLCCN     \undefined \def \showLCCN      #1{\unskip}     \fi
\ifx \shownote     \undefined \def \shownote      #1{#1}          \fi
\ifx \showarticletitle \undefined \def \showarticletitle #1{#1}   \fi
\ifx \showURL      \undefined \def \showURL       {\relax}        \fi
\providecommand\bibfield[2]{#2}
\providecommand\bibinfo[2]{#2}
\providecommand\natexlab[1]{#1}
\providecommand\showeprint[2][]{arXiv:#2}

\bibitem[noa(2023)]%
        {noauthor_wga_2023}
 \bibinfo{year}{2023}\natexlab{}.
\newblock \bibinfo{title}{{WGA} {Negotiations} {Status} as of {May} 1, 2023}.
\newblock
\newblock
\urldef\tempurl%
\url{https://www.wgacontract2023.org/the-campaign/wga-negotiations-status-as-of-5-1-2023}
\showURL{%
\tempurl}


\bibitem[Agarwal and Karahanna(2000)]%
        {agarwal_time_2000}
\bibfield{author}{\bibinfo{person}{Ritu Agarwal} {and} \bibinfo{person}{Elena Karahanna}.} \bibinfo{year}{2000}\natexlab{}.
\newblock \showarticletitle{Time {Flies} {When} {You}'re {Having} {Fun}: {Cognitive} {Absorption} and {Beliefs} about {Information} {Technology} {Usage}}.
\newblock \bibinfo{journal}{\emph{MIS Quarterly}} \bibinfo{volume}{24}, \bibinfo{number}{4} (\bibinfo{year}{2000}), \bibinfo{pages}{665--694}.
\newblock
\showISSN{0276-7783}
\urldef\tempurl%
\url{https://doi.org/10.2307/3250951}
\showDOI{\tempurl}
\newblock
\shownote{Publisher: Management Information Systems Research Center, University of Minnesota}.


\bibitem[Aki~Tamashiro(2021)]%
        {aki_tamashiro_how_2021}
\bibfield{author}{\bibinfo{person}{Mariana Aki~Tamashiro}.} \bibinfo{year}{2021}\natexlab{}.
\newblock \showarticletitle{How do we teach {Emerging} {Technologies} in {K}-9 {Education}? {Using} design fiction and constructionist approaches to support the understanding of emerging technologies’ societal implications in formal {K}-9 education}. In \bibinfo{booktitle}{\emph{Proceedings of the 20th {Annual} {ACM} {Interaction} {Design} and {Children} {Conference}}} \emph{(\bibinfo{series}{{IDC} '21})}. \bibinfo{publisher}{Association for Computing Machinery}, \bibinfo{address}{New York, NY, USA}, \bibinfo{pages}{637--640}.
\newblock
\showISBNx{978-1-4503-8452-0}
\urldef\tempurl%
\url{https://doi.org/10.1145/3459990.3463402}
\showDOI{\tempurl}


\bibitem[Banchi and Bell(2008)]%
        {banchi_many_2008}
\bibfield{author}{\bibinfo{person}{Heather Banchi} {and} \bibinfo{person}{Randy Bell}.} \bibinfo{year}{2008}\natexlab{}.
\newblock \showarticletitle{The many levels of inquiry}.
\newblock \bibinfo{journal}{\emph{Science and children}} \bibinfo{volume}{46}, \bibinfo{number}{2} (\bibinfo{year}{2008}), \bibinfo{pages}{26}.
\newblock
\newblock
\shownote{Publisher: National Science Teachers Association}.


\bibitem[Bardzell et~al\mbox{.}(2015)]%
        {bardzell_immodest_2015}
\bibfield{author}{\bibinfo{person}{Jeffrey Bardzell}, \bibinfo{person}{Shaowen Bardzell}, {and} \bibinfo{person}{Lone Koefoed~Hansen}.} \bibinfo{year}{2015}\natexlab{}.
\newblock \showarticletitle{Immodest {Proposals}: {Research} {Through} {Design} and {Knowledge}}. In \bibinfo{booktitle}{\emph{Proceedings of the 33rd {Annual} {ACM} {Conference} on {Human} {Factors} in {Computing} {Systems}}} \emph{(\bibinfo{series}{{CHI} '15})}. \bibinfo{publisher}{ACM}, \bibinfo{address}{New York, NY, USA}, \bibinfo{pages}{2093--2102}.
\newblock
\showISBNx{978-1-4503-3145-6}
\urldef\tempurl%
\url{https://doi.org/10.1145/2702123.2702400}
\showDOI{\tempurl}


\bibitem[Bender et~al\mbox{.}(2021)]%
        {bender_dangers_2021}
\bibfield{author}{\bibinfo{person}{Emily~M. Bender}, \bibinfo{person}{Timnit Gebru}, \bibinfo{person}{Angelina McMillan-Major}, {and} \bibinfo{person}{Shmargaret Shmitchell}.} \bibinfo{year}{2021}\natexlab{}.
\newblock \showarticletitle{On the {Dangers} of {Stochastic} {Parrots}: {Can} {Language} {Models} {Be} {Too} {Big}?}. In \bibinfo{booktitle}{\emph{Proceedings of the 2021 {ACM} {Conference} on {Fairness}, {Accountability}, and {Transparency}}} \emph{(\bibinfo{series}{{FAccT} '21})}. \bibinfo{publisher}{Association for Computing Machinery}, \bibinfo{address}{New York, NY, USA}, \bibinfo{pages}{610--623}.
\newblock
\showISBNx{978-1-4503-8309-7}
\urldef\tempurl%
\url{https://doi.org/10.1145/3442188.3445922}
\showDOI{\tempurl}


\bibitem[Benjamin et~al\mbox{.}(2021)]%
        {benjamin_machine_2021}
\bibfield{author}{\bibinfo{person}{Jesse~Josua Benjamin}, \bibinfo{person}{Arne Berger}, \bibinfo{person}{Nick Merrill}, {and} \bibinfo{person}{James Pierce}.} \bibinfo{year}{2021}\natexlab{}.
\newblock \showarticletitle{Machine {Learning} {Uncertainty} as a {Design} {Material}: {A} {Post}-{Phenomenological} {Inquiry}}. In \bibinfo{booktitle}{\emph{Proceedings of the 2021 {CHI} {Conference} on {Human} {Factors} in {Computing} {Systems}}} \emph{(\bibinfo{series}{{CHI} '21})}. \bibinfo{publisher}{Association for Computing Machinery}, \bibinfo{address}{New York, NY, USA}, \bibinfo{pages}{1--14}.
\newblock
\showISBNx{978-1-4503-8096-6}
\urldef\tempurl%
\url{https://doi.org/10.1145/3411764.3445481}
\showDOI{\tempurl}


\bibitem[Benjamin et~al\mbox{.}(2023)]%
        {benjamin_entoptic_2023}
\bibfield{author}{\bibinfo{person}{Jesse~Josua Benjamin}, \bibinfo{person}{Heidi Biggs}, \bibinfo{person}{Arne Berger}, \bibinfo{person}{Julija Rukanskaitė}, \bibinfo{person}{Michael~B. Heidt}, \bibinfo{person}{Nick Merrill}, \bibinfo{person}{James Pierce}, {and} \bibinfo{person}{Joseph Lindley}.} \bibinfo{year}{2023}\natexlab{}.
\newblock \showarticletitle{The {Entoptic} {Field} {Camera} as {Metaphor}-{Driven} {Research}-through-{Design} with {AI} {Technologies}}. In \bibinfo{booktitle}{\emph{Proceedings of the 2023 {CHI} {Conference} on {Human} {Factors} in {Computing} {Systems}}} \emph{(\bibinfo{series}{{CHI} '23})}. \bibinfo{publisher}{Association for Computing Machinery}, \bibinfo{address}{New York, NY, USA}, \bibinfo{pages}{1--19}.
\newblock
\showISBNx{978-1-4503-9421-5}
\urldef\tempurl%
\url{https://doi.org/10.1145/3544548.3581175}
\showDOI{\tempurl}


\bibitem[Benjamin et~al\mbox{.}(2022)]%
        {benjamin_explanation_2022}
\bibfield{author}{\bibinfo{person}{Jesse~Josua Benjamin}, \bibinfo{person}{Christoph Kinkeldey}, \bibinfo{person}{Claudia Müller-Birn}, \bibinfo{person}{Tim Korjakow}, {and} \bibinfo{person}{Eva-Maria Herbst}.} \bibinfo{year}{2022}\natexlab{}.
\newblock \showarticletitle{Explanation {Strategies} as an {Empirical}-{Analytical} {Lens} for {Socio}-{Technical} {Contextualization} of {Machine} {Learning} {Interpretability}}.
\newblock \bibinfo{journal}{\emph{Proceedings of the ACM on Human-Computer Interaction}} \bibinfo{volume}{6}, \bibinfo{number}{GROUP} (\bibinfo{date}{Jan.} \bibinfo{year}{2022}), \bibinfo{pages}{39:1--39:25}.
\newblock
\urldef\tempurl%
\url{https://doi.org/10.1145/3492858}
\showDOI{\tempurl}


\bibitem[Bilstrup et~al\mbox{.}(2022)]%
        {bilstrup_opportunities_2022}
\bibfield{author}{\bibinfo{person}{Karl-Emil~Kjær Bilstrup}, \bibinfo{person}{Magnus~Høholt Kaspersen}, \bibinfo{person}{Marie-Louise Stisen~Kjerstein Sørensen}, {and} \bibinfo{person}{Marianne~Graves Petersen}.} \bibinfo{year}{2022}\natexlab{}.
\newblock \showarticletitle{Opportunities and {Challenges} of {Teaching} {Machine} {Learning} as a {Design} {Material} with the micro:bit}. In \bibinfo{booktitle}{\emph{Adjunct {Proceedings} of the 2022 {Nordic} {Human}-{Computer} {Interaction} {Conference}}} \emph{(\bibinfo{series}{{NordiCHI} '22})}. \bibinfo{publisher}{Association for Computing Machinery}, \bibinfo{address}{New York, NY, USA}, \bibinfo{pages}{1--6}.
\newblock
\showISBNx{978-1-4503-9448-2}
\urldef\tempurl%
\url{https://doi.org/10.1145/3547522.3547689}
\showDOI{\tempurl}


\bibitem[Binns et~al\mbox{.}(2018)]%
        {binns_its_2018}
\bibfield{author}{\bibinfo{person}{Reuben Binns}, \bibinfo{person}{Max Van~Kleek}, \bibinfo{person}{Michael Veale}, \bibinfo{person}{Ulrik Lyngs}, \bibinfo{person}{Jun Zhao}, {and} \bibinfo{person}{Nigel Shadbolt}.} \bibinfo{year}{2018}\natexlab{}.
\newblock \showarticletitle{'{It}'s {Reducing} a {Human} {Being} to a {Percentage}': {Perceptions} of {Justice} in {Algorithmic} {Decisions}}. In \bibinfo{booktitle}{\emph{Proceedings of the 2018 {CHI} {Conference} on {Human} {Factors} in {Computing} {Systems}}} \emph{(\bibinfo{series}{{CHI} '18})}. \bibinfo{publisher}{Association for Computing Machinery}, \bibinfo{address}{New York, NY, USA}, \bibinfo{pages}{1--14}.
\newblock
\showISBNx{978-1-4503-5620-6}
\urldef\tempurl%
\url{https://doi.org/10.1145/3173574.3173951}
\showDOI{\tempurl}


\bibitem[Birhane et~al\mbox{.}(2021)]%
        {birhane_multimodal_2021}
\bibfield{author}{\bibinfo{person}{Abeba Birhane}, \bibinfo{person}{Vinay~Uday Prabhu}, {and} \bibinfo{person}{Emmanuel Kahembwe}.} \bibinfo{year}{2021}\natexlab{}.
\newblock \bibinfo{title}{Multimodal datasets: misogyny, pornography, and malignant stereotypes}.
\newblock
\newblock
\urldef\tempurl%
\url{https://doi.org/10.48550/arXiv.2110.01963}
\showDOI{\tempurl}
\newblock
\shownote{arXiv:2110.01963 [cs]}.


\bibitem[Bitmead(2023)]%
        {bitmead_problem_2023}
\bibfield{author}{\bibinfo{person}{Charlotte Bitmead}.} \bibinfo{year}{2023}\natexlab{}.
\newblock \bibinfo{title}{The problem with {TikTok}’s ‘{Bold} {Glamour}’ filter}.
\newblock
\newblock
\urldef\tempurl%
\url{https://www.cosmopolitan.com/uk/beauty-hair/beauty-trends/a43177683/tiktok-bold-glamour-filter/}
\showURL{%
\tempurl}
\newblock
\shownote{Section: Beauty Trends}.


\bibitem[Blikstein(2013)]%
        {blikstein_gears_2013}
\bibfield{author}{\bibinfo{person}{Paulo Blikstein}.} \bibinfo{year}{2013}\natexlab{}.
\newblock \showarticletitle{Gears of our childhood: constructionist toolkits, robotics, and physical computing, past and future}. In \bibinfo{booktitle}{\emph{Proceedings of the 12th {International} {Conference} on {Interaction} {Design} and {Children}}} \emph{(\bibinfo{series}{{IDC} '13})}. \bibinfo{publisher}{Association for Computing Machinery}, \bibinfo{address}{New York, NY, USA}, \bibinfo{pages}{173--182}.
\newblock
\showISBNx{978-1-4503-1918-8}
\urldef\tempurl%
\url{https://doi.org/10.1145/2485760.2485786}
\showDOI{\tempurl}


\bibitem[Bogost(2020)]%
        {bogost_your_2020}
\bibfield{author}{\bibinfo{person}{Ian Bogost}.} \bibinfo{year}{2020}\natexlab{}.
\newblock \bibinfo{title}{Your {Phone} {Wasn}’t {Built} for the {Apocalypse}}.
\newblock
\newblock
\urldef\tempurl%
\url{https://www.theatlantic.com/technology/archive/2020/09/camera-phone-wildfire-sky/616279/}
\showURL{%
\tempurl}
\newblock
\shownote{Section: Technology}.


\bibitem[Boucher(2023)]%
        {boucher_research_2023}
\bibfield{author}{\bibinfo{person}{Andy Boucher}.} \bibinfo{year}{2023}\natexlab{}.
\newblock \showarticletitle{Research {Products} at {Scale}: {Learnings} from {Designing} {Devices} in {Multiples} of {Ones}, {Tens}, {Hundreds} and {Thousands}}. In \bibinfo{booktitle}{\emph{Proceedings of the 2023 {CHI} {Conference} on {Human} {Factors} in {Computing} {Systems}}} \emph{(\bibinfo{series}{{CHI} '23})}. \bibinfo{publisher}{Association for Computing Machinery}, \bibinfo{address}{New York, NY, USA}, \bibinfo{pages}{1--15}.
\newblock
\showISBNx{978-1-4503-9421-5}
\urldef\tempurl%
\url{https://doi.org/10.1145/3544548.3581540}
\showDOI{\tempurl}


\bibitem[Chameau et~al\mbox{.}(2014)]%
        {chameau_foundational_2014}
\bibfield{author}{\bibinfo{person}{Jean-Lou Chameau}, \bibinfo{person}{William~F. Ballhaus}, {and} \bibinfo{person}{Herbert~S. Lin}.} \bibinfo{year}{2014}\natexlab{}.
\newblock \showarticletitle{Foundational {Technologies}}.
\newblock In \bibinfo{booktitle}{\emph{Emerging and {Readily} {Available} {Technologies} and {National} {Security}: {A} {Framework} for {Addressing} {Ethical}, {Legal}, and {Societal} {Issues}}}. \bibinfo{publisher}{National Academies Press (US)}.
\newblock
\urldef\tempurl%
\url{https://www.ncbi.nlm.nih.gov/books/NBK216326/}
\showURL{%
\tempurl}


\bibitem[{Culture, Media and Sport Committee}(2023)]%
        {culture_media_and_sport_committee_connected_2023}
\bibfield{author}{\bibinfo{person}{{Culture, Media and Sport Committee}}.} \bibinfo{year}{2023}\natexlab{}.
\newblock \bibinfo{booktitle}{\emph{Connected tech: {AI} and creative technology}}.
\newblock \bibinfo{type}{{T}echnical {R}eport} HC 1643. \bibinfo{institution}{House of Commons}, \bibinfo{address}{London, UK}.
\newblock
\urldef\tempurl%
\url{https://publications.parliament.uk/pa/cm5803/cmselect/cmcumeds/1643/report.html}
\showURL{%
\tempurl}


\bibitem[Dangol and Dasgupta(2023)]%
        {dangol_constructionist_2023}
\bibfield{author}{\bibinfo{person}{Aayushi Dangol} {and} \bibinfo{person}{Sayamindu Dasgupta}.} \bibinfo{year}{2023}\natexlab{}.
\newblock \showarticletitle{Constructionist approaches to critical data literacy: {A} review}. In \bibinfo{booktitle}{\emph{Proceedings of the 22nd {Annual} {ACM} {Interaction} {Design} and {Children} {Conference}}} \emph{(\bibinfo{series}{{IDC} '23})}. \bibinfo{publisher}{Association for Computing Machinery}, \bibinfo{address}{New York, NY, USA}, \bibinfo{pages}{112--123}.
\newblock
\showISBNx{9798400701313}
\urldef\tempurl%
\url{https://doi.org/10.1145/3585088.3589367}
\showDOI{\tempurl}


\bibitem[Debono(2021)]%
        {debono_collecting_2021}
\bibfield{author}{\bibinfo{person}{Sandro Debono}.} \bibinfo{year}{2021}\natexlab{}.
\newblock \showarticletitle{Collecting {Pandemic} {Phenomena}: {Reflections} on {Rapid} {Response} {Collecting} and the {Art} {Museum}}.
\newblock \bibinfo{journal}{\emph{Collections}} \bibinfo{volume}{17}, \bibinfo{number}{2} (\bibinfo{date}{June} \bibinfo{year}{2021}), \bibinfo{pages}{179--185}.
\newblock
\showISSN{1550-1906}
\urldef\tempurl%
\url{https://doi.org/10.1177/1550190620980844}
\showDOI{\tempurl}
\newblock
\shownote{Publisher: SAGE Publications Inc}.


\bibitem[{Department for Education}(2023)]%
        {department_for_education_generative_2023}
\bibfield{author}{\bibinfo{person}{{Department for Education}}.} \bibinfo{year}{2023}\natexlab{}.
\newblock \bibinfo{booktitle}{\emph{Generative artificial intelligence in education}}.
\newblock \bibinfo{type}{{T}echnical {R}eport}. \bibinfo{institution}{UK Government}.
\newblock
\urldef\tempurl%
\url{https://www.gov.uk/government/publications/generative-artificial-intelligence-in-education}
\showURL{%
\tempurl}


\bibitem[Dove and Fayard(2020)]%
        {dove_monsters_2020}
\bibfield{author}{\bibinfo{person}{Graham Dove} {and} \bibinfo{person}{Anne-Laure Fayard}.} \bibinfo{year}{2020}\natexlab{}.
\newblock \showarticletitle{Monsters, {Metaphors}, and {Machine} {Learning}}. In \bibinfo{booktitle}{\emph{Proceedings of the 2020 {CHI} {Conference} on {Human} {Factors} in {Computing} {Systems}}} \emph{(\bibinfo{series}{{CHI} '20})}. \bibinfo{publisher}{Association for Computing Machinery}, \bibinfo{address}{Honolulu, HI, USA}, \bibinfo{pages}{1--17}.
\newblock
\showISBNx{978-1-4503-6708-0}
\urldef\tempurl%
\url{https://doi.org/10.1145/3313831.3376275}
\showDOI{\tempurl}


\bibitem[Druga et~al\mbox{.}(2019)]%
        {druga_inclusive_2019}
\bibfield{author}{\bibinfo{person}{Stefania Druga}, \bibinfo{person}{Sarah~T. Vu}, \bibinfo{person}{Eesh Likhith}, {and} \bibinfo{person}{Tammy Qiu}.} \bibinfo{year}{2019}\natexlab{}.
\newblock \showarticletitle{Inclusive {AI} literacy for kids around the world}. In \bibinfo{booktitle}{\emph{Proceedings of {FabLearn} 2019}} \emph{(\bibinfo{series}{{FL2019}})}. \bibinfo{publisher}{Association for Computing Machinery}, \bibinfo{address}{New York, NY, USA}, \bibinfo{pages}{104--111}.
\newblock
\showISBNx{978-1-4503-6244-3}
\urldef\tempurl%
\url{https://doi.org/10.1145/3311890.3311904}
\showDOI{\tempurl}


\bibitem[Ehsan and Riedl(2020)]%
        {ehsan_human-centered_2020}
\bibfield{author}{\bibinfo{person}{Upol Ehsan} {and} \bibinfo{person}{Mark~O. Riedl}.} \bibinfo{year}{2020}\natexlab{}.
\newblock \showarticletitle{Human-{Centered} {Explainable} {AI}: {Towards} a {Reflective} {Sociotechnical} {Approach}}. In \bibinfo{booktitle}{\emph{{HCI} {International} 2020 - {Late} {Breaking} {Papers}: {Multimodality} and {Intelligence}}} \emph{(\bibinfo{series}{Lecture {Notes} in {Computer} {Science}})}, \bibfield{editor}{\bibinfo{person}{Constantine Stephanidis}, \bibinfo{person}{Masaaki Kurosu}, \bibinfo{person}{Helmut Degen}, {and} \bibinfo{person}{Lauren Reinerman-Jones}} (Eds.). \bibinfo{publisher}{Springer International Publishing}, \bibinfo{address}{Cham}, \bibinfo{pages}{449--466}.
\newblock
\showISBNx{978-3-030-60117-1}
\urldef\tempurl%
\url{https://doi.org/10.1007/978-3-030-60117-1_33}
\showDOI{\tempurl}


\bibitem[Emsley(2023)]%
        {emsley_chatgpt_2023}
\bibfield{author}{\bibinfo{person}{Robin Emsley}.} \bibinfo{year}{2023}\natexlab{}.
\newblock \showarticletitle{{ChatGPT}: these are not hallucinations – they’re fabrications and falsifications}.
\newblock \bibinfo{journal}{\emph{Schizophrenia}} \bibinfo{volume}{9}, \bibinfo{number}{1} (\bibinfo{date}{Aug.} \bibinfo{year}{2023}), \bibinfo{pages}{1--2}.
\newblock
\showISSN{2754-6993}
\urldef\tempurl%
\url{https://doi.org/10.1038/s41537-023-00379-4}
\showDOI{\tempurl}
\newblock
\shownote{Number: 1 Publisher: Nature Publishing Group}.


\bibitem[Etherington(2013)]%
        {etherington_interview_2013}
\bibfield{author}{\bibinfo{person}{Rose Etherington}.} \bibinfo{year}{2013}\natexlab{}.
\newblock \bibinfo{title}{Interview with {Kieran} {Long} on rapid response collecting at the {V}\&{A}}.
\newblock
\newblock
\urldef\tempurl%
\url{https://www.dezeen.com/2013/12/18/rapid-response-collecting-victoria-and-albert-museum-kieran-long/}
\showURL{%
\tempurl}
\newblock
\shownote{Section: all}.


\bibitem[Eynon and Young(2021)]%
        {eynon_methodology_2021}
\bibfield{author}{\bibinfo{person}{Rebecca Eynon} {and} \bibinfo{person}{Erin Young}.} \bibinfo{year}{2021}\natexlab{}.
\newblock \showarticletitle{Methodology, {Legend}, and {Rhetoric}: {The} {Constructions} of {AI} by {Academia}, {Industry}, and {Policy} {Groups} for {Lifelong} {Learning}}.
\newblock \bibinfo{journal}{\emph{Science, Technology, \& Human Values}} \bibinfo{volume}{46}, \bibinfo{number}{1} (\bibinfo{date}{Jan.} \bibinfo{year}{2021}), \bibinfo{pages}{166--191}.
\newblock
\showISSN{0162-2439}
\urldef\tempurl%
\url{https://doi.org/10.1177/0162243920906475}
\showDOI{\tempurl}
\newblock
\shownote{Publisher: SAGE Publications Inc}.


\bibitem[Fitzgerald(1994)]%
        {fitzgerald_theories_1994}
\bibfield{author}{\bibinfo{person}{M.~A. Fitzgerald}.} \bibinfo{year}{1994}\natexlab{}.
\newblock \showarticletitle{Theories of {Reflection} for {Learning}}.
\newblock In \bibinfo{booktitle}{\emph{Reflective {Practice} in {Nursing}: {The} {Growth} of the {Professional} {Practitioner}}}. \bibinfo{publisher}{Blackwell Scientific Publications}, \bibinfo{pages}{63--84}.
\newblock
\showISBNx{978-0-632-03597-7}
\urldef\tempurl%
\url{https://figshare.utas.edu.au/articles/chapter/Theories_of_Reflection_for_Learning/23054621/1}
\showURL{%
\tempurl}


\bibitem[for Education(2013)]%
        {noauthor_national_2013}
\bibfield{author}{\bibinfo{person}{Department for Education}.} \bibinfo{year}{2013}\natexlab{}.
\newblock \bibinfo{booktitle}{\emph{The national curriculum in {England}: {Key} stages 1 and 2 framework document}}.
\newblock \bibinfo{type}{{T}echnical {R}eport} DFE-00178-2013. \bibinfo{institution}{Department for Education}.
\newblock


\bibitem[Gaver(2012)]%
        {gaver_what_2012}
\bibfield{author}{\bibinfo{person}{William Gaver}.} \bibinfo{year}{2012}\natexlab{}.
\newblock \showarticletitle{What {Should} {We} {Expect} from {Research} {Through} {Design}?}. In \bibinfo{booktitle}{\emph{Proceedings of the {SIGCHI} {Conference} on {Human} {Factors} in {Computing} {Systems}}} \emph{(\bibinfo{series}{{CHI} '12})}. \bibinfo{publisher}{ACM}, \bibinfo{address}{New York, NY, USA}, \bibinfo{pages}{937--946}.
\newblock
\showISBNx{978-1-4503-1015-4}
\urldef\tempurl%
\url{https://doi.org/10.1145/2207676.2208538}
\showDOI{\tempurl}
\newblock
\shownote{event-place: Austin, Texas, USA}.


\bibitem[Gaver et~al\mbox{.}(2022)]%
        {gaver_emergence_2022}
\bibfield{author}{\bibinfo{person}{William Gaver}, \bibinfo{person}{Peter~Gall Krogh}, \bibinfo{person}{Andy Boucher}, {and} \bibinfo{person}{David Chatting}.} \bibinfo{year}{2022}\natexlab{}.
\newblock \showarticletitle{Emergence as a {Feature} of {Practice}-based {Design} {Research}}. In \bibinfo{booktitle}{\emph{Designing {Interactive} {Systems} {Conference}}} \emph{(\bibinfo{series}{{DIS} '22})}. \bibinfo{publisher}{Association for Computing Machinery}, \bibinfo{address}{New York, NY, USA}, \bibinfo{pages}{517--526}.
\newblock
\showISBNx{978-1-4503-9358-4}
\urldef\tempurl%
\url{https://doi.org/10.1145/3532106.3533524}
\showDOI{\tempurl}


\bibitem[Gaver et~al\mbox{.}(2003)]%
        {gaver_ambiguity_2003}
\bibfield{author}{\bibinfo{person}{William~W. Gaver}, \bibinfo{person}{Jacob Beaver}, {and} \bibinfo{person}{Steve Benford}.} \bibinfo{year}{2003}\natexlab{}.
\newblock \showarticletitle{Ambiguity as a {Resource} for {Design}}. In \bibinfo{booktitle}{\emph{Proceedings of {CHI} 2003}}. \bibinfo{publisher}{ACM Press}, \bibinfo{pages}{269--270}.
\newblock


\bibitem[Gaver et~al\mbox{.}(2004)]%
        {gaver_drift_2004}
\bibfield{author}{\bibinfo{person}{William~W. Gaver}, \bibinfo{person}{John Bowers}, \bibinfo{person}{Andrew Boucher}, \bibinfo{person}{Hans Gellerson}, \bibinfo{person}{Sarah Pennington}, \bibinfo{person}{Albrecht Schmidt}, \bibinfo{person}{Anthony Steed}, \bibinfo{person}{Nicholas Villars}, {and} \bibinfo{person}{Brendan Walker}.} \bibinfo{year}{2004}\natexlab{}.
\newblock \showarticletitle{The {Drift} {Table}: {Designing} for {Ludic} {Engagement}}. In \bibinfo{booktitle}{\emph{{CHI} '04 {Extended} {Abstracts} on {Human} {Factors} in {Computing} {Systems}}} \emph{(\bibinfo{series}{{CHI} {EA} '04})}. \bibinfo{publisher}{ACM}, \bibinfo{address}{New York, NY, USA}, \bibinfo{pages}{885--900}.
\newblock
\showISBNx{978-1-58113-703-3}
\urldef\tempurl%
\url{https://doi.org/10.1145/985921.985947}
\showDOI{\tempurl}
\newblock
\shownote{event-place: Vienna, Austria}.


\bibitem[Giannini(2023)]%
        {giannini_generative_2023}
\bibfield{author}{\bibinfo{person}{Stefania Giannini}.} \bibinfo{year}{2023}\natexlab{}.
\newblock \bibinfo{booktitle}{\emph{Generative {AI} and the future of education}}.
\newblock \bibinfo{type}{{T}echnical {R}eport} ED/ADG/2023/02. \bibinfo{institution}{UNESCO}. \bibinfo{pages}{8} pages.
\newblock


\bibitem[Gulson and Witzenberger(2022)]%
        {gulson_repackaging_2022}
\bibfield{author}{\bibinfo{person}{Kalervo~N. Gulson} {and} \bibinfo{person}{Kevin Witzenberger}.} \bibinfo{year}{2022}\natexlab{}.
\newblock \showarticletitle{Repackaging authority: artificial intelligence, automated governance and education trade shows}.
\newblock \bibinfo{journal}{\emph{Journal of Education Policy}} \bibinfo{volume}{37}, \bibinfo{number}{1} (\bibinfo{date}{Jan.} \bibinfo{year}{2022}), \bibinfo{pages}{145--160}.
\newblock
\showISSN{0268-0939}
\urldef\tempurl%
\url{https://doi.org/10.1080/02680939.2020.1785552}
\showDOI{\tempurl}
\newblock
\shownote{Publisher: Routledge \_eprint: https://doi.org/10.1080/02680939.2020.1785552}.


\bibitem[Halskov and Hansen(2015)]%
        {halskov_diversity_2015}
\bibfield{author}{\bibinfo{person}{Kim Halskov} {and} \bibinfo{person}{Nicolai~Brodersen Hansen}.} \bibinfo{year}{2015}\natexlab{}.
\newblock \showarticletitle{The diversity of participatory design research practice at {PDC} 2002–2012}.
\newblock \bibinfo{journal}{\emph{International Journal of Human-Computer Studies}}  \bibinfo{volume}{74} (\bibinfo{date}{Feb.} \bibinfo{year}{2015}), \bibinfo{pages}{81--92}.
\newblock
\showISSN{1071-5819}
\urldef\tempurl%
\url{https://doi.org/10.1016/j.ijhcs.2014.09.003}
\showDOI{\tempurl}


\bibitem[Han and Cai(2023)]%
        {han_design_2023}
\bibfield{author}{\bibinfo{person}{Ariel Han} {and} \bibinfo{person}{Zhenyao Cai}.} \bibinfo{year}{2023}\natexlab{}.
\newblock \showarticletitle{Design implications of generative {AI} systems for visual storytelling for young learners}. In \bibinfo{booktitle}{\emph{Proceedings of the 22nd {Annual} {ACM} {Interaction} {Design} and {Children} {Conference}}} \emph{(\bibinfo{series}{{IDC} '23})}. \bibinfo{publisher}{Association for Computing Machinery}, \bibinfo{address}{New York, NY, USA}, \bibinfo{pages}{470--474}.
\newblock
\showISBNx{9798400701313}
\urldef\tempurl%
\url{https://doi.org/10.1145/3585088.3593867}
\showDOI{\tempurl}


\bibitem[Hautea et~al\mbox{.}(2017)]%
        {hautea_youth_2017}
\bibfield{author}{\bibinfo{person}{Samantha Hautea}, \bibinfo{person}{Sayamindu Dasgupta}, {and} \bibinfo{person}{Benjamin~Mako Hill}.} \bibinfo{year}{2017}\natexlab{}.
\newblock \showarticletitle{Youth {Perspectives} on {Critical} {Data} {Literacies}}. In \bibinfo{booktitle}{\emph{Proceedings of the 2017 {CHI} {Conference} on {Human} {Factors} in {Computing} {Systems}}} \emph{(\bibinfo{series}{{CHI} '17})}. \bibinfo{publisher}{Association for Computing Machinery}, \bibinfo{address}{New York, NY, USA}, \bibinfo{pages}{919--930}.
\newblock
\showISBNx{978-1-4503-4655-9}
\urldef\tempurl%
\url{https://doi.org/10.1145/3025453.3025823}
\showDOI{\tempurl}


\bibitem[Hohman et~al\mbox{.}(2019)]%
        {hohman_gamut:_2019}
\bibfield{author}{\bibinfo{person}{Fred Hohman}, \bibinfo{person}{Andrew Head}, \bibinfo{person}{Rich Caruana}, \bibinfo{person}{Robert DeLine}, {and} \bibinfo{person}{Steven~M. Drucker}.} \bibinfo{year}{2019}\natexlab{}.
\newblock \showarticletitle{Gamut: {A} {Design} {Probe} to {Understand} {How} {Data} {Scientists} {Understand} {Machine} {Learning} {Models}}. In \bibinfo{booktitle}{\emph{Proceedings of the 2019 {CHI} {Conference} on {Human} {Factors} in {Computing} {Systems}}} \emph{(\bibinfo{series}{{CHI} '19})}. \bibinfo{publisher}{ACM}, \bibinfo{address}{New York, NY, USA}, \bibinfo{pages}{579:1--579:13}.
\newblock
\showISBNx{978-1-4503-5970-2}
\urldef\tempurl%
\url{https://doi.org/10.1145/3290605.3300809}
\showDOI{\tempurl}
\newblock
\shownote{event-place: Glasgow, Scotland Uk}.


\bibitem[Hu(2023)]%
        {hu_chatgpt_2023}
\bibfield{author}{\bibinfo{person}{Krystal Hu}.} \bibinfo{year}{2023}\natexlab{}.
\newblock \showarticletitle{{ChatGPT} sets record for fastest-growing user base - analyst note}.
\newblock \bibinfo{journal}{\emph{Reuters}} (\bibinfo{date}{Feb.} \bibinfo{year}{2023}).
\newblock
\urldef\tempurl%
\url{https://www.reuters.com/technology/chatgpt-sets-record-fastest-growing-user-base-analyst-note-2023-02-01/}
\showURL{%
\tempurl}


\bibitem[Höök and Löwgren(2012)]%
        {hook_strong_2012}
\bibfield{author}{\bibinfo{person}{Kristina Höök} {and} \bibinfo{person}{Jonas Löwgren}.} \bibinfo{year}{2012}\natexlab{}.
\newblock \showarticletitle{Strong concepts: {Intermediate}-level knowledge in interaction design research}.
\newblock \bibinfo{journal}{\emph{ACM Transactions on Computer-Human Interaction}} \bibinfo{volume}{19}, \bibinfo{number}{3} (\bibinfo{date}{Oct.} \bibinfo{year}{2012}), \bibinfo{pages}{23:1--23:18}.
\newblock
\showISSN{1073-0516}
\urldef\tempurl%
\url{https://doi.org/10.1145/2362364.2362371}
\showDOI{\tempurl}


\bibitem[Ingold(2018)]%
        {ingold_back_2018}
\bibfield{author}{\bibinfo{person}{Tim Ingold}.} \bibinfo{year}{2018}\natexlab{}.
\newblock \showarticletitle{Back to the future with the theory of affordances}.
\newblock \bibinfo{journal}{\emph{HAU: Journal of Ethnographic Theory}} \bibinfo{volume}{8}, \bibinfo{number}{1-2} (\bibinfo{date}{March} \bibinfo{year}{2018}), \bibinfo{pages}{39--44}.
\newblock
\showISSN{2575-1433}
\urldef\tempurl%
\url{https://doi.org/10.1086/698358}
\showDOI{\tempurl}
\newblock
\shownote{Publisher: The University of Chicago Press}.


\bibitem[Jiang et~al\mbox{.}(2022)]%
        {jiang_promptmaker_2022}
\bibfield{author}{\bibinfo{person}{Ellen Jiang}, \bibinfo{person}{Kristen Olson}, \bibinfo{person}{Edwin Toh}, \bibinfo{person}{Alejandra Molina}, \bibinfo{person}{Aaron Donsbach}, \bibinfo{person}{Michael Terry}, {and} \bibinfo{person}{Carrie~J Cai}.} \bibinfo{year}{2022}\natexlab{}.
\newblock \showarticletitle{{PromptMaker}: {Prompt}-based {Prototyping} with {Large} {Language} {Models}}. In \bibinfo{booktitle}{\emph{Extended {Abstracts} of the 2022 {CHI} {Conference} on {Human} {Factors} in {Computing} {Systems}}} \emph{(\bibinfo{series}{{CHI} {EA} '22})}. \bibinfo{publisher}{Association for Computing Machinery}, \bibinfo{address}{New York, NY, USA}, \bibinfo{pages}{1--8}.
\newblock
\showISBNx{978-1-4503-9156-6}
\urldef\tempurl%
\url{https://doi.org/10.1145/3491101.3503564}
\showDOI{\tempurl}


\bibitem[Judd(2020)]%
        {judd_all_2020}
\bibfield{author}{\bibinfo{person}{Sarah Judd}.} \bibinfo{year}{2020}\natexlab{}.
\newblock \showarticletitle{All {Means} {All}: {Bringing} {Project}-based, {Approachable} {AI} {Curriculum} to {More} {High} {School} {Students} through {AI4ALL} {Open} {Learning}}. In \bibinfo{booktitle}{\emph{Proceedings of the 51st {ACM} {Technical} {Symposium} on {Computer} {Science} {Education}}} \emph{(\bibinfo{series}{{SIGCSE} '20})}. \bibinfo{publisher}{Association for Computing Machinery}, \bibinfo{address}{New York, NY, USA}, \bibinfo{pages}{1409}.
\newblock
\showISBNx{978-1-4503-6793-6}
\urldef\tempurl%
\url{https://doi.org/10.1145/3328778.3372554}
\showDOI{\tempurl}


\bibitem[Kafai(2005)]%
        {sawyer_constructionism_2005}
\bibfield{author}{\bibinfo{person}{Yasmin~B. Kafai}.} \bibinfo{year}{2005}\natexlab{}.
\newblock \showarticletitle{Constructionism}.
\newblock In \bibinfo{booktitle}{\emph{The {Cambridge} {Handbook} of the {Learning} {Sciences}}}, \bibfield{editor}{\bibinfo{person}{R.~Keith Sawyer}} (Ed.). \bibinfo{publisher}{Cambridge University Press}, \bibinfo{address}{Cambridge, UNITED KINGDOM}, \bibinfo{pages}{35--46}.
\newblock
\showISBNx{978-0-511-21910-8}
\urldef\tempurl%
\url{http://ebookcentral.proquest.com/lib/lancaster/detail.action?docID=261112}
\showURL{%
\tempurl}


\bibitem[Kafai and Burke(2015)]%
        {kafai_constructionist_2015}
\bibfield{author}{\bibinfo{person}{Yasmin~B. Kafai} {and} \bibinfo{person}{Quinn Burke}.} \bibinfo{year}{2015}\natexlab{}.
\newblock \showarticletitle{Constructionist {Gaming}: {Understanding} the {Benefits} of {Making} {Games} for {Learning}}.
\newblock \bibinfo{journal}{\emph{Educational Psychologist}} \bibinfo{volume}{50}, \bibinfo{number}{4} (\bibinfo{date}{Oct.} \bibinfo{year}{2015}), \bibinfo{pages}{313--334}.
\newblock
\showISSN{0046-1520}
\urldef\tempurl%
\url{https://doi.org/10.1080/00461520.2015.1124022}
\showDOI{\tempurl}


\bibitem[Kaur et~al\mbox{.}(2020)]%
        {kaur_interpreting_2020}
\bibfield{author}{\bibinfo{person}{Harmanpreet Kaur}, \bibinfo{person}{Harsha Nori}, \bibinfo{person}{Samuel Jenkins}, \bibinfo{person}{Rich Caruana}, \bibinfo{person}{Hanna Wallach}, {and} \bibinfo{person}{Jennifer Wortman~Vaughan}.} \bibinfo{year}{2020}\natexlab{}.
\newblock \showarticletitle{Interpreting {Interpretability}: {Understanding} {Data} {Scientists}' {Use} of {Interpretability} {Tools} for {Machine} {Learning}}. In \bibinfo{booktitle}{\emph{Proceedings of the 2020 {CHI} {Conference} on {Human} {Factors} in {Computing} {Systems}}} \emph{(\bibinfo{series}{{CHI} '20})}. \bibinfo{publisher}{Association for Computing Machinery}, \bibinfo{address}{Honolulu, HI, USA}, \bibinfo{pages}{1--14}.
\newblock
\showISBNx{978-1-4503-6708-0}
\urldef\tempurl%
\url{https://doi.org/10.1145/3313831.3376219}
\showDOI{\tempurl}


\bibitem[Klein(2023)]%
        {klein_chatgpt_2023}
\bibfield{author}{\bibinfo{person}{Alyson Klein}.} \bibinfo{year}{2023}\natexlab{}.
\newblock \showarticletitle{{ChatGPT} {Cheating}: {What} to {Do} {When} {It} {Happens}}.
\newblock \bibinfo{journal}{\emph{Education Week}} (\bibinfo{date}{Feb.} \bibinfo{year}{2023}).
\newblock
\showISSN{0277-4232}
\urldef\tempurl%
\url{https://www.edweek.org/technology/chatgpt-cheating-what-to-do-when-it-happens/2023/02}
\showURL{%
\tempurl}


\bibitem[Kolb et~al\mbox{.}(2001)]%
        {kolb_experiential_2001}
\bibfield{author}{\bibinfo{person}{David~A. Kolb}, \bibinfo{person}{Richard~E. Boyatzis}, {and} \bibinfo{person}{Charalampos Mainemelis}.} \bibinfo{year}{2001}\natexlab{}.
\newblock \showarticletitle{Experiential learning theory: {Previous} research and new directions}.
\newblock In \bibinfo{booktitle}{\emph{Perspectives on thinking, learning, and cognitive styles}}. \bibinfo{publisher}{Lawrence Erlbaum Associates Publishers}, \bibinfo{address}{Mahwah, NJ, US}, \bibinfo{pages}{227--247}.
\newblock
\showISBNx{978-0-8058-3430-7 978-0-8058-3431-4}


\bibitem[Krogh and Koskinen(2020)]%
        {krogh_drifting_2020}
\bibfield{author}{\bibinfo{person}{Peter~Gall Krogh} {and} \bibinfo{person}{Ilpo Koskinen}.} \bibinfo{year}{2020}\natexlab{}.
\newblock \bibinfo{booktitle}{\emph{Drifting by {Intention}: {Four} {Epistemic} {Traditions} from within {Constructive} {Design} {Research}}}.
\newblock \bibinfo{publisher}{Springer International Publishing}, \bibinfo{address}{Cham}.
\newblock
\showISBNx{978-3-030-37895-0 978-3-030-37896-7}
\urldef\tempurl%
\url{https://doi.org/10.1007/978-3-030-37896-7}
\showDOI{\tempurl}


\bibitem[Kudina and Verbeek(2019)]%
        {kudina_ethics_2019}
\bibfield{author}{\bibinfo{person}{Olya Kudina} {and} \bibinfo{person}{Peter-Paul Verbeek}.} \bibinfo{year}{2019}\natexlab{}.
\newblock \showarticletitle{Ethics from {Within}: {Google} {Glass}, the {Collingridge} {Dilemma}, and the {Mediated} {Value} of {Privacy}}.
\newblock \bibinfo{journal}{\emph{Science, Technology, \& Human Values}} \bibinfo{volume}{44}, \bibinfo{number}{2} (\bibinfo{date}{March} \bibinfo{year}{2019}), \bibinfo{pages}{291--314}.
\newblock
\showISSN{0162-2439}
\urldef\tempurl%
\url{https://doi.org/10.1177/0162243918793711}
\showDOI{\tempurl}


\bibitem[Lau(1997)]%
        {lau_review_1997}
\bibfield{author}{\bibinfo{person}{F. Lau}.} \bibinfo{year}{1997}\natexlab{}.
\newblock \showarticletitle{A {Review} on the {Use} of {Action} {Research} in {Information} {Systems} {Studies}}.
\newblock In \bibinfo{booktitle}{\emph{Information {Systems} and {Qualitative} {Research}: {Proceedings} of the {IFIP} {TC8} {WG} 8.2 {International} {Conference} on {Information} {Systems} and {Qualitative} {Research}, 31st {May}–3rd {June} 1997, {Philadelphia}, {Pennsylvania}, {USA}}}, \bibfield{editor}{\bibinfo{person}{Allen~S. Lee}, \bibinfo{person}{Jonathan Liebenau}, {and} \bibinfo{person}{Janice~I. DeGross}} (Eds.). \bibinfo{publisher}{Springer US}, \bibinfo{address}{Boston, MA}, \bibinfo{pages}{31--68}.
\newblock
\showISBNx{978-0-387-35309-8}
\urldef\tempurl%
\url{https://doi.org/10.1007/978-0-387-35309-8_4}
\showDOI{\tempurl}


\bibitem[Lee et~al\mbox{.}(2021)]%
        {lee_developing_2021}
\bibfield{author}{\bibinfo{person}{Irene Lee}, \bibinfo{person}{Safinah Ali}, \bibinfo{person}{Helen Zhang}, \bibinfo{person}{Daniella DiPaola}, {and} \bibinfo{person}{Cynthia Breazeal}.} \bibinfo{year}{2021}\natexlab{}.
\newblock \showarticletitle{Developing {Middle} {School} {Students}' {AI} {Literacy}}. In \bibinfo{booktitle}{\emph{Proceedings of the 52nd {ACM} {Technical} {Symposium} on {Computer} {Science} {Education}}} \emph{(\bibinfo{series}{{SIGCSE} '21})}. \bibinfo{publisher}{Association for Computing Machinery}, \bibinfo{address}{New York, NY, USA}, \bibinfo{pages}{191--197}.
\newblock
\showISBNx{978-1-4503-8062-1}
\urldef\tempurl%
\url{https://doi.org/10.1145/3408877.3432513}
\showDOI{\tempurl}


\bibitem[Lee et~al\mbox{.}(2023)]%
        {lee_prompt_2023}
\bibfield{author}{\bibinfo{person}{Unggi Lee}, \bibinfo{person}{Ariel Han}, \bibinfo{person}{Jeongjin Lee}, \bibinfo{person}{Eunseo Lee}, \bibinfo{person}{Jiwon Kim}, \bibinfo{person}{Hyeoncheol Kim}, {and} \bibinfo{person}{Cheolil Lim}.} \bibinfo{year}{2023}\natexlab{}.
\newblock \showarticletitle{Prompt {Aloud}!: {Incorporating} image-generative {AI} into {STEAM} class with learning analytics using prompt data}.
\newblock \bibinfo{journal}{\emph{Education and Information Technologies}} (\bibinfo{date}{Sept.} \bibinfo{year}{2023}).
\newblock
\showISSN{1573-7608}
\urldef\tempurl%
\url{https://doi.org/10.1007/s10639-023-12150-4}
\showDOI{\tempurl}


\bibitem[Lee and Jung(2020)]%
        {lee_ludic-hri_2020}
\bibfield{author}{\bibinfo{person}{Wen-Ying Lee} {and} \bibinfo{person}{Malte Jung}.} \bibinfo{year}{2020}\natexlab{}.
\newblock \showarticletitle{Ludic-{HRI}: {Designing} {Playful} {Experiences} with {Robots}}. In \bibinfo{booktitle}{\emph{Companion of the 2020 {ACM}/{IEEE} {International} {Conference} on {Human}-{Robot} {Interaction}}} \emph{(\bibinfo{series}{{HRI} '20})}. \bibinfo{publisher}{Association for Computing Machinery}, \bibinfo{address}{New York, NY, USA}, \bibinfo{pages}{582--584}.
\newblock
\showISBNx{978-1-4503-7057-8}
\urldef\tempurl%
\url{https://doi.org/10.1145/3371382.3377429}
\showDOI{\tempurl}


\bibitem[Levin and Tsybulsky(2017)]%
        {levin_constructionist_2017}
\bibfield{author}{\bibinfo{person}{Ilya Levin} {and} \bibinfo{person}{Dina Tsybulsky}.} \bibinfo{year}{2017}\natexlab{}.
\newblock \showarticletitle{The {Constructionist} {Learning} {Approach} in the {Digital} {Age}}.
\newblock \bibinfo{journal}{\emph{Creative Education}} \bibinfo{volume}{8}, \bibinfo{number}{15} (\bibinfo{date}{Dec.} \bibinfo{year}{2017}), \bibinfo{pages}{2463--2475}.
\newblock
\urldef\tempurl%
\url{https://doi.org/10.4236/ce.2017.815169}
\showDOI{\tempurl}
\newblock
\shownote{Number: 15 Publisher: Scientific Research Publishing}.


\bibitem[Liao et~al\mbox{.}(2020)]%
        {liao_questioning_2020}
\bibfield{author}{\bibinfo{person}{Q.~Vera Liao}, \bibinfo{person}{Daniel Gruen}, {and} \bibinfo{person}{Sarah Miller}.} \bibinfo{year}{2020}\natexlab{}.
\newblock \showarticletitle{Questioning the {AI}: {Informing} {Design} {Practices} for {Explainable} {AI} {User} {Experiences}}. In \bibinfo{booktitle}{\emph{Proceedings of the 2020 {CHI} {Conference} on {Human} {Factors} in {Computing} {Systems}}} \emph{(\bibinfo{series}{{CHI} '20})}. \bibinfo{publisher}{Association for Computing Machinery}, \bibinfo{address}{New York, NY, USA}, \bibinfo{pages}{1--15}.
\newblock
\showISBNx{978-1-4503-6708-0}
\urldef\tempurl%
\url{https://doi.org/10.1145/3313831.3376590}
\showDOI{\tempurl}


\bibitem[Lindley et~al\mbox{.}(2020)]%
        {lindley_researching_2020}
\bibfield{author}{\bibinfo{person}{Joseph Lindley}, \bibinfo{person}{Haider~Ali Akmal}, \bibinfo{person}{Franziska Pillling}, {and} \bibinfo{person}{Paul Coulton}.} \bibinfo{year}{2020}\natexlab{}.
\newblock \showarticletitle{Researching {AI} {Legibility} through {Design}}. In \bibinfo{booktitle}{\emph{Proceedings of the 2020 {CHI} {Conference} on {Human} {Factors} in {Computing} {Systems}}} \emph{(\bibinfo{series}{{CHI} '20})}. \bibinfo{publisher}{Association for Computing Machinery}, \bibinfo{address}{Honolulu, HI, USA}, \bibinfo{pages}{1--13}.
\newblock
\showISBNx{978-1-4503-6708-0}
\urldef\tempurl%
\url{https://doi.org/10.1145/3313831.3376792}
\showDOI{\tempurl}


\bibitem[Lindley et~al\mbox{.}(2017)]%
        {lindley_implications_2017}
\bibfield{author}{\bibinfo{person}{Joseph Lindley}, \bibinfo{person}{Paul Coulton}, {and} \bibinfo{person}{Miriam Sturdee}.} \bibinfo{year}{2017}\natexlab{}.
\newblock \showarticletitle{Implications for {Adoption}}. In \bibinfo{booktitle}{\emph{Proceedings of the 2017 {CHI} {Conference} on {Human} {Factors} in {Computing} {Systems}}} \emph{(\bibinfo{series}{{CHI} '17})}. \bibinfo{publisher}{Association for Computing Machinery}, \bibinfo{address}{New York, NY, USA}, \bibinfo{pages}{265--277}.
\newblock
\showISBNx{978-1-4503-4655-9}
\urldef\tempurl%
\url{https://doi.org/10.1145/3025453.3025742}
\showDOI{\tempurl}


\bibitem[Long and Magerko(2020)]%
        {long_what_2020}
\bibfield{author}{\bibinfo{person}{Duri Long} {and} \bibinfo{person}{Brian Magerko}.} \bibinfo{year}{2020}\natexlab{}.
\newblock \showarticletitle{What is {AI} {Literacy}? {Competencies} and {Design} {Considerations}}. In \bibinfo{booktitle}{\emph{Proceedings of the 2020 {CHI} {Conference} on {Human} {Factors} in {Computing} {Systems}}} \emph{(\bibinfo{series}{{CHI} '20})}. \bibinfo{publisher}{Association for Computing Machinery}, \bibinfo{address}{New York, NY, USA}, \bibinfo{pages}{1--16}.
\newblock
\showISBNx{978-1-4503-6708-0}
\urldef\tempurl%
\url{https://doi.org/10.1145/3313831.3376727}
\showDOI{\tempurl}


\bibitem[Madaio et~al\mbox{.}(2020)]%
        {madaio_co-designing_2020}
\bibfield{author}{\bibinfo{person}{Michael~A. Madaio}, \bibinfo{person}{Luke Stark}, \bibinfo{person}{Jennifer Wortman~Vaughan}, {and} \bibinfo{person}{Hanna Wallach}.} \bibinfo{year}{2020}\natexlab{}.
\newblock \showarticletitle{Co-{Designing} {Checklists} to {Understand} {Organizational} {Challenges} and {Opportunities} around {Fairness} in {AI}}. In \bibinfo{booktitle}{\emph{Proceedings of the 2020 {CHI} {Conference} on {Human} {Factors} in {Computing} {Systems}}} \emph{(\bibinfo{series}{{CHI} '20})}. \bibinfo{publisher}{Association for Computing Machinery}, \bibinfo{address}{Honolulu, HI, USA}, \bibinfo{pages}{1--14}.
\newblock
\showISBNx{978-1-4503-6708-0}
\urldef\tempurl%
\url{https://doi.org/10.1145/3313831.3376445}
\showDOI{\tempurl}


\bibitem[Maloney et~al\mbox{.}(2010)]%
        {maloney_scratch_2010}
\bibfield{author}{\bibinfo{person}{John Maloney}, \bibinfo{person}{Mitchel Resnick}, \bibinfo{person}{Natalie Rusk}, \bibinfo{person}{Brian Silverman}, {and} \bibinfo{person}{Evelyn Eastmond}.} \bibinfo{year}{2010}\natexlab{}.
\newblock \showarticletitle{The {Scratch} {Programming} {Language} and {Environment}}.
\newblock \bibinfo{journal}{\emph{ACM Transactions on Computing Education}} \bibinfo{volume}{10}, \bibinfo{number}{4} (\bibinfo{date}{Nov.} \bibinfo{year}{2010}), \bibinfo{pages}{16:1--16:15}.
\newblock
\urldef\tempurl%
\url{https://doi.org/10.1145/1868358.1868363}
\showDOI{\tempurl}


\bibitem[Mirnig et~al\mbox{.}(2020)]%
        {mirnig_blinded_2020}
\bibfield{author}{\bibinfo{person}{Alexander~G. Mirnig}, \bibinfo{person}{Magdalena Gärtner}, \bibinfo{person}{Alexander Meschtscherjakov}, {and} \bibinfo{person}{Manfred Tscheligi}.} \bibinfo{year}{2020}\natexlab{}.
\newblock \showarticletitle{Blinded by novelty: a reflection on participant curiosity and novelty in automated vehicle studies based on experiences from the field}. In \bibinfo{booktitle}{\emph{Proceedings of {Mensch} und {Computer} 2020}} \emph{(\bibinfo{series}{{MuC} '20})}. \bibinfo{publisher}{Association for Computing Machinery}, \bibinfo{address}{New York, NY, USA}, \bibinfo{pages}{373--381}.
\newblock
\showISBNx{978-1-4503-7540-5}
\urldef\tempurl%
\url{https://doi.org/10.1145/3404983.3405593}
\showDOI{\tempurl}


\bibitem[Mirsky and Lee(2021)]%
        {mirsky_creation_2021}
\bibfield{author}{\bibinfo{person}{Yisroel Mirsky} {and} \bibinfo{person}{Wenke Lee}.} \bibinfo{year}{2021}\natexlab{}.
\newblock \showarticletitle{The {Creation} and {Detection} of {Deepfakes}: {A} {Survey}}.
\newblock \bibinfo{journal}{\emph{Comput. Surveys}} \bibinfo{volume}{54}, \bibinfo{number}{1} (\bibinfo{date}{Jan.} \bibinfo{year}{2021}), \bibinfo{pages}{7:1--7:41}.
\newblock
\showISSN{0360-0300}
\urldef\tempurl%
\url{https://doi.org/10.1145/3425780}
\showDOI{\tempurl}


\bibitem[Nemorin et~al\mbox{.}(2023)]%
        {nemorin_ai_2023}
\bibfield{author}{\bibinfo{person}{Selena Nemorin}, \bibinfo{person}{Andreas Vlachidis}, \bibinfo{person}{Hayford~M. Ayerakwa}, {and} \bibinfo{person}{Panagiotis Andriotis}.} \bibinfo{year}{2023}\natexlab{}.
\newblock \showarticletitle{{AI} hyped? {A} horizon scan of discourse on artificial intelligence in education ({AIED}) and development}.
\newblock \bibinfo{journal}{\emph{Learning, Media and Technology}} \bibinfo{volume}{48}, \bibinfo{number}{1} (\bibinfo{date}{Jan.} \bibinfo{year}{2023}), \bibinfo{pages}{38--51}.
\newblock
\showISSN{1743-9884}
\urldef\tempurl%
\url{https://doi.org/10.1080/17439884.2022.2095568}
\showDOI{\tempurl}
\newblock
\shownote{Publisher: Routledge \_eprint: https://doi.org/10.1080/17439884.2022.2095568}.


\bibitem[Nerantzi et~al\mbox{.}(2023)]%
        {nerantzi_101_2023}
\bibfield{author}{\bibinfo{person}{Chrissi Nerantzi}, \bibinfo{person}{Sandra Abegglen}, \bibinfo{person}{Marianna Karatsiori}, {and} \bibinfo{person}{Antonio Martínez-Arboleda~(Eds.)}.} \bibinfo{year}{2023}\natexlab{}.
\newblock \bibinfo{booktitle}{\emph{101 creative ideas to use {AI} in education, {A} crowdsourced collection}}.
\newblock \bibinfo{publisher}{Zenodo}.
\newblock
\urldef\tempurl%
\url{https://doi.org/10.5281/zenodo.8072950}
\showDOI{\tempurl}
\newblock
\shownote{Version Number: 2023 1.0}.


\bibitem[Norman(1999)]%
        {norman_affordance_1999}
\bibfield{author}{\bibinfo{person}{Donald~A. Norman}.} \bibinfo{year}{1999}\natexlab{}.
\newblock \showarticletitle{Affordance, conventions, and design}.
\newblock \bibinfo{journal}{\emph{interactions}} \bibinfo{volume}{6}, \bibinfo{number}{3} (\bibinfo{year}{1999}), \bibinfo{pages}{38--43}.
\newblock
\newblock
\shownote{Publisher: ACM New York, NY, USA}.


\bibitem[Odom et~al\mbox{.}(2016)]%
        {odom_research_2016}
\bibfield{author}{\bibinfo{person}{William Odom}, \bibinfo{person}{Ron Wakkary}, \bibinfo{person}{Youn-kyung Lim}, \bibinfo{person}{Audrey Desjardins}, \bibinfo{person}{Bart Hengeveld}, {and} \bibinfo{person}{Richard Banks}.} \bibinfo{year}{2016}\natexlab{}.
\newblock \showarticletitle{From {Research} {Prototype} to {Research} {Product}}. In \bibinfo{booktitle}{\emph{Proceedings of the 2016 {CHI} {Conference} on {Human} {Factors} in {Computing} {Systems}}} \emph{(\bibinfo{series}{{CHI} '16})}. \bibinfo{publisher}{ACM}, \bibinfo{address}{New York, NY, USA}, \bibinfo{pages}{2549--2561}.
\newblock
\showISBNx{978-1-4503-3362-7}
\urldef\tempurl%
\url{https://doi.org/10.1145/2858036.2858447}
\showDOI{\tempurl}


\bibitem[Papert(1991)]%
        {papert_situating_1991}
\bibfield{author}{\bibinfo{person}{Seymour Papert}.} \bibinfo{year}{1991}\natexlab{}.
\newblock \showarticletitle{Situating constructionism}.
\newblock In \bibinfo{booktitle}{\emph{Constructionism}}, \bibfield{editor}{\bibinfo{person}{Seymour Papert} {and} \bibinfo{person}{Idit Harel}} (Eds.). Vol.~\bibinfo{volume}{36}. \bibinfo{publisher}{Ablex Publishing Corporation}, \bibinfo{pages}{1--11}.
\newblock
\urldef\tempurl%
\url{https://pirun.ku.ac.th/~btun/papert/sitcons.pdf}
\showURL{%
\tempurl}


\bibitem[Perrotta and Selwyn(2020)]%
        {perrotta_deep_2020}
\bibfield{author}{\bibinfo{person}{Carlo Perrotta} {and} \bibinfo{person}{Neil Selwyn}.} \bibinfo{year}{2020}\natexlab{}.
\newblock \showarticletitle{Deep learning goes to school: toward a relational understanding of {AI} in education}.
\newblock \bibinfo{journal}{\emph{Learning, Media and Technology}} \bibinfo{volume}{45}, \bibinfo{number}{3} (\bibinfo{date}{July} \bibinfo{year}{2020}), \bibinfo{pages}{251--269}.
\newblock
\showISSN{1743-9884}
\urldef\tempurl%
\url{https://doi.org/10.1080/17439884.2020.1686017}
\showDOI{\tempurl}
\newblock
\shownote{Publisher: Routledge \_eprint: https://doi.org/10.1080/17439884.2020.1686017}.


\bibitem[{Pew Research Center}(2020)]%
        {pew_research_center_childrens_2020}
\bibfield{author}{\bibinfo{person}{{Pew Research Center}}.} \bibinfo{year}{2020}\natexlab{}.
\newblock \bibinfo{title}{Children’s engagement with digital devices, screen time}.
\newblock
\newblock
\urldef\tempurl%
\url{https://www.pewresearch.org/internet/2020/07/28/childrens-engagement-with-digital-devices-screen-time/}
\showURL{%
\tempurl}


\bibitem[Pittarello et~al\mbox{.}(2017)]%
        {pittarello_hci_2017}
\bibfield{author}{\bibinfo{person}{Fabio Pittarello}, \bibinfo{person}{Gualtiero Volpe}, {and} \bibinfo{person}{Massimo Zancanaro}.} \bibinfo{year}{2017}\natexlab{}.
\newblock \showarticletitle{{HCI} and education in a changing world: from school to public engagement}. In \bibinfo{booktitle}{\emph{Proceedings of the 12th {Biannual} {Conference} on {Italian} {SIGCHI} {Chapter}}} \emph{(\bibinfo{series}{{CHItaly} '17})}. \bibinfo{publisher}{Association for Computing Machinery}, \bibinfo{address}{New York, NY, USA}, \bibinfo{pages}{1--2}.
\newblock
\showISBNx{978-1-4503-5237-6}
\urldef\tempurl%
\url{https://doi.org/10.1145/3125571.3125576}
\showDOI{\tempurl}


\bibitem[Redström(2017)]%
        {redstrom_making_2017}
\bibfield{author}{\bibinfo{person}{Johan Redström}.} \bibinfo{year}{2017}\natexlab{}.
\newblock \bibinfo{booktitle}{\emph{Making {Design} {Theory}}}.
\newblock \bibinfo{publisher}{The MIT Press}.
\newblock
\showISBNx{978-0-262-34184-4}
\urldef\tempurl%
\url{https://doi.org/10.7551/mitpress/11160.001.0001}
\showDOI{\tempurl}


\bibitem[Rombach et~al\mbox{.}(2022)]%
        {rombach_high-resolution_2022}
\bibfield{author}{\bibinfo{person}{Robin Rombach}, \bibinfo{person}{Andreas Blattmann}, \bibinfo{person}{Dominik Lorenz}, \bibinfo{person}{Patrick Esser}, {and} \bibinfo{person}{Björn Ommer}.} \bibinfo{year}{2022}\natexlab{}.
\newblock \bibinfo{title}{High-{Resolution} {Image} {Synthesis} with {Latent} {Diffusion} {Models}}.
\newblock
\newblock
\urldef\tempurl%
\url{https://doi.org/10.48550/arXiv.2112.10752}
\showDOI{\tempurl}
\newblock
\shownote{arXiv:2112.10752 [cs]}.


\bibitem[Rubegni et~al\mbox{.}(2022)]%
        {rubegni_dont_2022}
\bibfield{author}{\bibinfo{person}{Elisa Rubegni}, \bibinfo{person}{Laura Malinverni}, {and} \bibinfo{person}{Jason Yip}.} \bibinfo{year}{2022}\natexlab{}.
\newblock \showarticletitle{“{Don}'t let the robots walk our dogs, but it's ok for them to do our homework”: children's perceptions, fears, and hopes in social robots.}. In \bibinfo{booktitle}{\emph{Proceedings of the 21st {Annual} {ACM} {Interaction} {Design} and {Children} {Conference}}} \emph{(\bibinfo{series}{{IDC} '22})}. \bibinfo{publisher}{Association for Computing Machinery}, \bibinfo{address}{New York, NY, USA}, \bibinfo{pages}{352--361}.
\newblock
\showISBNx{978-1-4503-9197-9}
\urldef\tempurl%
\url{https://doi.org/10.1145/3501712.3529726}
\showDOI{\tempurl}


\bibitem[Saha et~al\mbox{.}(2020)]%
        {saha_measuring_2020}
\bibfield{author}{\bibinfo{person}{Debjani Saha}, \bibinfo{person}{Candice Schumann}, \bibinfo{person}{Duncan~C. McElfresh}, \bibinfo{person}{John~P. Dickerson}, \bibinfo{person}{Michelle~L. Mazurek}, {and} \bibinfo{person}{Michael~Carl Tschantz}.} \bibinfo{year}{2020}\natexlab{}.
\newblock \showarticletitle{Measuring {Non}-{Expert} {Comprehension} of {Machine} {Learning} {Fairness} {Metrics}}. In \bibinfo{booktitle}{\emph{{PMLR}}}, Vol.~\bibinfo{volume}{119}.
\newblock
\urldef\tempurl%
\url{http://arxiv.org/abs/2001.00089}
\showURL{%
\tempurl}
\newblock
\shownote{arXiv: 2001.00089}.


\bibitem[Satariano and Kang(2023)]%
        {satariano_how_2023}
\bibfield{author}{\bibinfo{person}{Adam Satariano} {and} \bibinfo{person}{Cecilia Kang}.} \bibinfo{year}{2023}\natexlab{}.
\newblock \showarticletitle{How {Nations} {Are} {Losing} a {Global} {Race} to {Tackle} {A}.{I}.’s {Harms}}.
\newblock \bibinfo{journal}{\emph{The New York Times}} (\bibinfo{date}{Dec.} \bibinfo{year}{2023}).
\newblock
\urldef\tempurl%
\url{https://www.nytimes.com/2023/12/06/technology/ai-regulation-policies.html}
\showURL{%
\tempurl}


\bibitem[Schaper et~al\mbox{.}(2022)]%
        {schaper_computational_2022}
\bibfield{author}{\bibinfo{person}{Marie-Monique Schaper}, \bibinfo{person}{Rachel~Charlotte Smith}, \bibinfo{person}{Mariana~Aki Tamashiro}, \bibinfo{person}{Maarten Van~Mechelen}, \bibinfo{person}{Mille~Skovhus Lunding}, \bibinfo{person}{Karl-Emil~Kjæer Bilstrup}, \bibinfo{person}{Magnus~Høholt Kaspersen}, \bibinfo{person}{Kasper~Løvborg Jensen}, \bibinfo{person}{Marianne~Graves Petersen}, {and} \bibinfo{person}{Ole~Sejer Iversen}.} \bibinfo{year}{2022}\natexlab{}.
\newblock \showarticletitle{Computational empowerment in practice: {Scaffolding} teenagers’ learning about emerging technologies and their ethical and societal impact}.
\newblock \bibinfo{journal}{\emph{International Journal of Child-Computer Interaction}} \bibinfo{volume}{34}, \bibinfo{number}{C} (\bibinfo{date}{Dec.} \bibinfo{year}{2022}).
\newblock
\showISSN{2212-8689}
\urldef\tempurl%
\url{https://doi.org/10.1016/j.ijcci.2022.100537}
\showDOI{\tempurl}


\bibitem[Schuler and Namioka(1993)]%
        {schuler_participatory_1993}
\bibfield{author}{\bibinfo{person}{Douglas Schuler} {and} \bibinfo{person}{Aki Namioka}.} \bibinfo{year}{1993}\natexlab{}.
\newblock \bibinfo{booktitle}{\emph{Participatory {Design}: {Principles} and {Practices}}}.
\newblock \bibinfo{publisher}{CRC Press}.
\newblock
\showISBNx{978-0-8058-0951-0}
\newblock
\shownote{Google-Books-ID: pWOEk6Sk4YkC}.


\bibitem[Schön(1983)]%
        {schon_reflective_1983}
\bibfield{author}{\bibinfo{person}{Donald~A. Schön}.} \bibinfo{year}{1983}\natexlab{}.
\newblock \bibinfo{booktitle}{\emph{The reflective practitioner: how professionals think in action}}.
\newblock \bibinfo{publisher}{Basic Books}, \bibinfo{address}{New York}.
\newblock
\showISBNx{978-0-465-06878-4 978-0-465-06874-6}


\bibitem[Sengers and Gaver(2006)]%
        {sengers_staying_2006}
\bibfield{author}{\bibinfo{person}{Phoebe Sengers} {and} \bibinfo{person}{Bill Gaver}.} \bibinfo{year}{2006}\natexlab{}.
\newblock \showarticletitle{Staying {Open} to {Interpretation}: {Engaging} {Multiple} {Meanings} in {Design} and {Evaluation}}. In \bibinfo{booktitle}{\emph{Proceedings of the 6th {Conference} on {Designing} {Interactive} {Systems}}} \emph{(\bibinfo{series}{{DIS} '06})}. \bibinfo{publisher}{ACM}, \bibinfo{address}{New York, NY, USA}, \bibinfo{pages}{99--108}.
\newblock
\showISBNx{978-1-59593-367-6}
\urldef\tempurl%
\url{https://doi.org/10.1145/1142405.1142422}
\showDOI{\tempurl}


\bibitem[Smith et~al\mbox{.}(2023)]%
        {smith_research_2023}
\bibfield{author}{\bibinfo{person}{Rachel~Charlotte Smith}, \bibinfo{person}{Marie-Monique Schaper}, \bibinfo{person}{Mariana~Aki Tamashiro}, \bibinfo{person}{Maarten Van~Mechelen}, \bibinfo{person}{Marianne~Graves Petersen}, {and} \bibinfo{person}{Ole~Sejer Iversen}.} \bibinfo{year}{2023}\natexlab{}.
\newblock \showarticletitle{A research agenda for computational empowerment for emerging technology education}.
\newblock \bibinfo{journal}{\emph{International Journal of Child-Computer Interaction}}  \bibinfo{volume}{38} (\bibinfo{date}{Dec.} \bibinfo{year}{2023}), \bibinfo{pages}{100616}.
\newblock
\showISSN{2212-8689}
\urldef\tempurl%
\url{https://doi.org/10.1016/j.ijcci.2023.100616}
\showDOI{\tempurl}


\bibitem[Springer et~al\mbox{.}(2017)]%
        {springer_dice_2017}
\bibfield{author}{\bibinfo{person}{Aaron Springer}, \bibinfo{person}{Victoria Hollis}, {and} \bibinfo{person}{Steve Whittaker}.} \bibinfo{year}{2017}\natexlab{}.
\newblock \showarticletitle{Dice in the black box: {User} experiences with an inscrutable algorithm}. In \bibinfo{booktitle}{\emph{The 2017 {AAAI} {Spring} {Symposium} {Series}}}.
\newblock


\bibitem[Stappers and Giaccardi(2019)]%
        {stappers_research_2019}
\bibfield{author}{\bibinfo{person}{Pieter Stappers} {and} \bibinfo{person}{Elisa Giaccardi}.} \bibinfo{year}{2019}\natexlab{}.
\newblock \showarticletitle{Research through {Design}}.
\newblock In \bibinfo{booktitle}{\emph{The {Encyclopedia} of {Interaction} {Design}} (\bibinfo{edition}{2} ed.)}. \bibinfo{publisher}{Interaction Design Foundation}.
\newblock
\urldef\tempurl%
\url{https://www.interaction-design.org/literature/book/the-encyclopedia-of-human-computer-interaction-2nd-ed/research-through-design}
\showURL{%
\tempurl}


\bibitem[Su and Zhong(2022)]%
        {su_artificial_2022}
\bibfield{author}{\bibinfo{person}{Jiahong Su} {and} \bibinfo{person}{Yuchun Zhong}.} \bibinfo{year}{2022}\natexlab{}.
\newblock \showarticletitle{Artificial {Intelligence} ({AI}) in early childhood education: {Curriculum} design and future directions}.
\newblock \bibinfo{journal}{\emph{Computers and Education: Artificial Intelligence}}  \bibinfo{volume}{3} (\bibinfo{date}{Jan.} \bibinfo{year}{2022}), \bibinfo{pages}{100072}.
\newblock
\showISSN{2666-920X}
\urldef\tempurl%
\url{https://doi.org/10.1016/j.caeai.2022.100072}
\showDOI{\tempurl}


\bibitem[{UNESCO}(2022)]%
        {unesco_k-12_2022}
\bibfield{author}{\bibinfo{person}{{UNESCO}}.} \bibinfo{year}{2022}\natexlab{}.
\newblock \bibinfo{booktitle}{\emph{K-12 {AI} curricula: a mapping of government-endorsed {AI} curricula}}.
\newblock \bibinfo{type}{{T}echnical {R}eport} 0000380602. \bibinfo{institution}{UNESCO}, \bibinfo{address}{Paris}.
\newblock
\urldef\tempurl%
\url{https://unesdoc.unesco.org/ark:/48223/pf0000380602}
\showURL{%
\tempurl}


\bibitem[Van~Mechelen et~al\mbox{.}(2020)]%
        {van_mechelen_18_2020}
\bibfield{author}{\bibinfo{person}{Maarten Van~Mechelen}, \bibinfo{person}{Gökçe~Elif Baykal}, \bibinfo{person}{Christian Dindler}, \bibinfo{person}{Eva Eriksson}, {and} \bibinfo{person}{Ole~Sejer Iversen}.} \bibinfo{year}{2020}\natexlab{}.
\newblock \showarticletitle{18 {Years} of ethics in child-computer interaction research: a systematic literature review}. In \bibinfo{booktitle}{\emph{Proceedings of the {Interaction} {Design} and {Children} {Conference}}} \emph{(\bibinfo{series}{{IDC} '20})}. \bibinfo{publisher}{Association for Computing Machinery}, \bibinfo{address}{New York, NY, USA}, \bibinfo{pages}{161--183}.
\newblock
\showISBNx{978-1-4503-7981-6}
\urldef\tempurl%
\url{https://doi.org/10.1145/3392063.3394407}
\showDOI{\tempurl}


\bibitem[{Victoria and Albert Museum}(2017)]%
        {victoria_and_albert_museum_rapid_2017}
\bibfield{author}{\bibinfo{person}{{Victoria and Albert Museum}}.} \bibinfo{year}{2017}\natexlab{}.
\newblock \bibinfo{title}{Rapid {Response} {Collecting} {\textbar} {V}\&{A}}.
\newblock
\newblock
\urldef\tempurl%
\url{https://www.youtube.com/watch?v=_9AQXrKLkVY}
\showURL{%
\tempurl}


\bibitem[Walsh et~al\mbox{.}(2022)]%
        {walsh_making_2022}
\bibfield{author}{\bibinfo{person}{Benjamin Walsh}, \bibinfo{person}{Safinah Ali}, \bibinfo{person}{Francisco Castro}, \bibinfo{person}{Kayla Desportes}, \bibinfo{person}{Daniella DiPaola}, \bibinfo{person}{Irene Lee}, \bibinfo{person}{William Payne}, \bibinfo{person}{Scott Sieke}, {and} \bibinfo{person}{Helen Zhang}.} \bibinfo{year}{2022}\natexlab{}.
\newblock \showarticletitle{Making {Art} with and about {Artificial} {Intelligence}: {Three} {Approaches} to {Teaching} {AI} and {AI} {Ethics} to {Middle} and {High} {School} {Students}}. In \bibinfo{booktitle}{\emph{Proceedings of the 53rd {ACM} {Technical} {Symposium} on {Computer} {Science} {Education} {V}. 2}} \emph{(\bibinfo{series}{{SIGCSE} 2022})}. \bibinfo{publisher}{Association for Computing Machinery}, \bibinfo{address}{New York, NY, USA}, \bibinfo{pages}{1203}.
\newblock
\showISBNx{978-1-4503-9071-2}
\urldef\tempurl%
\url{https://doi.org/10.1145/3478432.3499157}
\showDOI{\tempurl}


\bibitem[Wang et~al\mbox{.}(2019)]%
        {wang_designing_2019}
\bibfield{author}{\bibinfo{person}{Danding Wang}, \bibinfo{person}{Qian Yang}, \bibinfo{person}{Ashraf Abdul}, {and} \bibinfo{person}{Brian~Y. Lim}.} \bibinfo{year}{2019}\natexlab{}.
\newblock \showarticletitle{Designing {Theory}-{Driven} {User}-{Centric} {Explainable} {AI}}. In \bibinfo{booktitle}{\emph{Proceedings of the 2019 {CHI} {Conference} on {Human} {Factors} in {Computing} {Systems}}} \emph{(\bibinfo{series}{{CHI} '19})}. \bibinfo{publisher}{Association for Computing Machinery}, \bibinfo{address}{Glasgow, Scotland Uk}, \bibinfo{pages}{1--15}.
\newblock
\showISBNx{978-1-4503-5970-2}
\urldef\tempurl%
\url{https://doi.org/10.1145/3290605.3300831}
\showDOI{\tempurl}


\bibitem[Wiggers(2023)]%
        {wiggers_stability_2023}
\bibfield{author}{\bibinfo{person}{Kyle Wiggers}.} \bibinfo{year}{2023}\natexlab{}.
\newblock \bibinfo{title}{Stability {AI} releases its latest image-generating model, {Stable} {Diffusion} {XL} 1.0}.
\newblock
\newblock
\urldef\tempurl%
\url{https://techcrunch.com/2023/07/26/stability-ai-releases-its-latest-image-generating-model-stable-diffusion-xl-1-0/}
\showURL{%
\tempurl}


\bibitem[Williams et~al\mbox{.}(2019)]%
        {williams_popbots_2019}
\bibfield{author}{\bibinfo{person}{Randi Williams}, \bibinfo{person}{Hae~Won Park}, \bibinfo{person}{Lauren Oh}, {and} \bibinfo{person}{Cynthia Breazeal}.} \bibinfo{year}{2019}\natexlab{}.
\newblock \showarticletitle{{PopBots}: {Designing} an {Artificial} {Intelligence} {Curriculum} for {Early} {Childhood} {Education}}.
\newblock \bibinfo{journal}{\emph{Proceedings of the AAAI Conference on Artificial Intelligence}} \bibinfo{volume}{33}, \bibinfo{number}{01} (\bibinfo{date}{July} \bibinfo{year}{2019}), \bibinfo{pages}{9729--9736}.
\newblock
\showISSN{2374-3468}
\urldef\tempurl%
\url{https://doi.org/10.1609/aaai.v33i01.33019729}
\showDOI{\tempurl}
\newblock
\shownote{Number: 01}.


\bibitem[Williamson(2023)]%
        {williamson_degenerative_2023}
\bibfield{author}{\bibinfo{person}{Ben Williamson}.} \bibinfo{year}{2023}\natexlab{}.
\newblock \bibinfo{title}{Degenerative {AI} in education}.
\newblock
\newblock
\urldef\tempurl%
\url{https://codeactsineducation.wordpress.com/2023/06/30/degenerative-ai-in-education/}
\showURL{%
\tempurl}


\bibitem[Williamson and Eynon(2020)]%
        {williamson_historical_2020}
\bibfield{author}{\bibinfo{person}{Ben Williamson} {and} \bibinfo{person}{Rebecca Eynon}.} \bibinfo{year}{2020}\natexlab{}.
\newblock \showarticletitle{Historical threads, missing links, and future directions in {AI} in education}.
\newblock \bibinfo{journal}{\emph{Learning, Media and Technology}} \bibinfo{volume}{45}, \bibinfo{number}{3} (\bibinfo{date}{July} \bibinfo{year}{2020}), \bibinfo{pages}{223--235}.
\newblock
\showISSN{1743-9884}
\urldef\tempurl%
\url{https://doi.org/10.1080/17439884.2020.1798995}
\showDOI{\tempurl}
\newblock
\shownote{Publisher: Routledge \_eprint: https://doi.org/10.1080/17439884.2020.1798995}.


\bibitem[Yannier et~al\mbox{.}(2021)]%
        {yannier_active_2021}
\bibfield{author}{\bibinfo{person}{Nesra Yannier}, \bibinfo{person}{Scott~E. Hudson}, \bibinfo{person}{Kenneth~R. Koedinger}, \bibinfo{person}{Kathy Hirsh-Pasek}, \bibinfo{person}{Roberta~Michnick Golinkoff}, \bibinfo{person}{Yuko Munakata}, \bibinfo{person}{Sabine Doebel}, \bibinfo{person}{Daniel~L. Schwartz}, \bibinfo{person}{Louis Deslauriers}, \bibinfo{person}{Logan McCarty}, \bibinfo{person}{Kristina Callaghan}, \bibinfo{person}{Elli~J. Theobald}, \bibinfo{person}{Scott Freeman}, \bibinfo{person}{Katelyn~M. Cooper}, {and} \bibinfo{person}{Sara~E. Brownell}.} \bibinfo{year}{2021}\natexlab{}.
\newblock \showarticletitle{Active learning: “{Hands}-on” meets “minds-on”}.
\newblock \bibinfo{journal}{\emph{Science}} \bibinfo{volume}{374}, \bibinfo{number}{6563} (\bibinfo{date}{Oct.} \bibinfo{year}{2021}), \bibinfo{pages}{26--30}.
\newblock
\urldef\tempurl%
\url{https://doi.org/10.1126/science.abj9957}
\showDOI{\tempurl}
\newblock
\shownote{Publisher: American Association for the Advancement of Science}.


\bibitem[Zembylas(2023)]%
        {zembylas_decolonial_2023}
\bibfield{author}{\bibinfo{person}{Michalinos Zembylas}.} \bibinfo{year}{2023}\natexlab{}.
\newblock \showarticletitle{A decolonial approach to {AI} in higher education teaching and learning: strategies for undoing the ethics of digital neocolonialism}.
\newblock \bibinfo{journal}{\emph{Learning, Media and Technology}} \bibinfo{volume}{48}, \bibinfo{number}{1} (\bibinfo{date}{Jan.} \bibinfo{year}{2023}), \bibinfo{pages}{25--37}.
\newblock
\showISSN{1743-9884}
\urldef\tempurl%
\url{https://doi.org/10.1080/17439884.2021.2010094}
\showDOI{\tempurl}
\newblock
\shownote{Publisher: Routledge \_eprint: https://doi.org/10.1080/17439884.2021.2010094}.


\bibitem[Zimmerman and Forlizzi(2014)]%
        {olson_research_2014}
\bibfield{author}{\bibinfo{person}{John Zimmerman} {and} \bibinfo{person}{Jodi Forlizzi}.} \bibinfo{year}{2014}\natexlab{}.
\newblock \showarticletitle{Research {Through} {Design} in {HCI}}.
\newblock In \bibinfo{booktitle}{\emph{Ways of {Knowing} in {HCI}}}, \bibfield{editor}{\bibinfo{person}{Judith~S. Olson} {and} \bibinfo{person}{Wendy~A. Kellogg}} (Eds.). \bibinfo{publisher}{Springer New York}, \bibinfo{address}{New York, NY}, \bibinfo{pages}{167--189}.
\newblock
\showISBNx{978-1-4939-0377-1 978-1-4939-0378-8}
\urldef\tempurl%
\url{https://doi.org/10.1007/978-1-4939-0378-8_8}
\showDOI{\tempurl}


\bibitem[Zimmerman et~al\mbox{.}(2007)]%
        {zimmerman_research_2007}
\bibfield{author}{\bibinfo{person}{John Zimmerman}, \bibinfo{person}{Jodi Forlizzi}, {and} \bibinfo{person}{Shelley Evenson}.} \bibinfo{year}{2007}\natexlab{}.
\newblock \showarticletitle{Research {Through} {Design} {As} a {Method} for {Interaction} {Design} {Research} in {HCI}}. In \bibinfo{booktitle}{\emph{Proceedings of the {SIGCHI} {Conference} on {Human} {Factors} in {Computing} {Systems}}} \emph{(\bibinfo{series}{{CHI} '07})}. \bibinfo{publisher}{ACM}, \bibinfo{address}{New York, NY, USA}, \bibinfo{pages}{493--502}.
\newblock
\showISBNx{978-1-59593-593-9}
\urldef\tempurl%
\url{https://doi.org/10.1145/1240624.1240704}
\showDOI{\tempurl}
\newblock
\shownote{event-place: San Jose, California, USA}.


\bibitem[Zimmerman et~al\mbox{.}(2010)]%
        {zimmerman_analysis_2010}
\bibfield{author}{\bibinfo{person}{John Zimmerman}, \bibinfo{person}{Erik Stolterman}, {and} \bibinfo{person}{Jodi Forlizzi}.} \bibinfo{year}{2010}\natexlab{}.
\newblock \showarticletitle{An {Analysis} and {Critique} of {Research} {Through} {Design}: {Towards} a {Formalization} of a {Research} {Approach}}. In \bibinfo{booktitle}{\emph{Proceedings of the 8th {ACM} {Conference} on {Designing} {Interactive} {Systems}}} \emph{(\bibinfo{series}{{DIS} '10})}. \bibinfo{publisher}{ACM}, \bibinfo{address}{New York, NY, USA}, \bibinfo{pages}{310--319}.
\newblock
\showISBNx{978-1-4503-0103-9}
\urldef\tempurl%
\url{https://doi.org/10.1145/1858171.1858228}
\showDOI{\tempurl}
\newblock
\shownote{event-place: Aarhus, Denmark}.


\end{thebibliography}

\appendix

\end{document}